\documentclass[]{aastex631}

\defcitealias{scope_paper1}{I}
\defcitealias{scope_paper2}{II}

\newcommand{\paperone}{{Paper \citetalias{scope_paper1}}}
\newcommand{\papertwo}{{Paper \citetalias{scope_paper2}}}

\newcommand{\scope}{\texttt{SCoPe}}

\usepackage{longtable}
\usepackage[flushleft]{threeparttable}
\usepackage{tabularx}
\usepackage{multirow}
\usepackage{graphicx}
\usepackage{amsmath,amssymb}
\usepackage{color}
\usepackage{units}
\usepackage{epstopdf}
\usepackage{hyperref}
\usepackage{multirow}
\usepackage{url}
\usepackage{subfigure}
\usepackage{rotating}
\usepackage{enumitem}\setlist[description]{font=\textendash\enskip\scshape\bfseries}
\usepackage{booktabs}

\newcommand{\beq}{\begin{equation}}
\newcommand{\eeq}{\end{equation}}
\newcommand{\bdm}{\begin{displaymath}}
\newcommand{\edm}{\end{displaymath}}
\newcommand{\T}[1]{\tilde{#1}}

\definecolor{Gray}{gray}{0.9}
\definecolor{orange}{rgb}{0.9,0.5,0}

\newcommand{\ztfink}[1]{ZTF-Fink}

\received{November 30, 2023}
\revised{February 22, 2024}
\accepted{March 12, 2024}

\begin{document}

\title{The ZTF Source Classification Project: III. A Catalog of Variable Sources}

\author[0000-0002-7718-7884]{Brian F. Healy}
\affil{School of Physics and Astronomy, University of Minnesota, Minneapolis, MN 55455, USA}

\author[0000-0002-8262-2924]{Michael W. Coughlin}
\affil{School of Physics and Astronomy, University of Minnesota, Minneapolis, MN 55455, USA}

\author[0000-0003-2242-0244]{Ashish A. Mahabal}
\affil{Division of Physics, Mathematics and Astronomy, California Institute of Technology, Pasadena, CA 91125, USA}
\affil{Center for Data Driven Discovery, California Institute of Technology, Pasadena, CA 91125, USA}

\author[0009-0003-6181-4526]{Theophile Jegou du Laz}
\affil{Division of Physics, Mathematics and Astronomy, California Institute of Technology, Pasadena, CA 91125, USA}

\author{Andrew Drake}
\affil{Division of Physics, Mathematics and Astronomy, California Institute of Technology, Pasadena, CA 91125, USA}

\author{Matthew J. Graham}
\affil{Division of Physics, Mathematics and Astronomy, California Institute of Technology, Pasadena, CA 91125, USA}

\author{Lynne A. Hillenbrand}
\affil{Department of Astronomy, California Institute of Technology, Pasadena, CA 91125, USA}

\author[0000-0002-2626-2872]{Jan van Roestel}
\affil{Anton Pannekoek Institute for Astronomy, University of Amsterdam, 1090 GE, Amsterdam, The Netherlands}

\author[0000-0003-4373-7777]{Paula Szkody}
\affil{Department of Astronomy, University of Washington, 
Seattle, WA 98195, USA}

\author{LeighAnna Zielske}
\affil{Department of Physics, Pacific Lutheran University, Tacoma, WA 98447, USA}
\affil{School of Physics and Astronomy, University of Minnesota, Minneapolis, MN 55455, USA}

\author{Mohammed Guiga}
\affil{Department of Computer Science and Engineering, University of Minnesota, Minneapolis, MN 55455, USA}

\author{Muhammad Yusuf Hassan}
\affil{Department of Electrical Engineering, Indian Institute of Technology Gandhinagar, Gujarat 382055, India}
\affil{Division of Physics, Mathematics and Astronomy, California Institute of Technology, Pasadena, CA 91125, USA}

\author{Jill L. Hughes}
\affil{School of Physics and Astronomy, University of Minnesota, Minneapolis, MN 55455, USA}

\author{Guy Nir}
\affil{Department of Astronomy, University of California, Berkeley, CA 94720, USA}
\affil{Lawrence Berkeley National Laboratory, 1 Cyclotron Road, MS 50B-4206, Berkeley, CA 94720, USA}

\author{Saagar Parikh}
\affil{Department of Electrical and Computer Engineering, Carnegie Mellon University, Pittsburgh, PA 15213, USA}
\affil{Computer Vision, Imaging and Graphics Lab, Indian Institute of Technology Gandhinagar, Gujarat 382055, India}

\author{Sungmin Park}
\affil{School of Statistics, University of Minnesota, Minneapolis, MN 55455, USA}

\author{Palak Purohit}
\affil{Division of Engineering and Applied Science, California Institute of Technology, Pasadena, CA 91125, USA}
\affil{Department of Electrical Engineering, Indian Institute of Technology Gandhinagar, Gujarat 382055, India}

\author{Umaa Rebbapragada}
\affil{Jet Propulsion Laboratory, California Institute of Technology, Pasadena, CA 91125, USA}

\author{Draco Reed}
\affil{School of Physics and Astronomy, University of Minnesota, Minneapolis, MN 55455, USA}

\author{Daniel Warshofsky}
\affil{School of Physics and Astronomy, University of Minnesota, Minneapolis, MN 55455, USA}

\author[0000-0002-9998-6732]{Avery Wold}
\affil{IPAC, California Institute of Technology, 1200 E. California Blvd., Pasadena, CA 91125, USA}

\author[0000-0002-7777-216X]{Joshua S. Bloom}
\affil{Department of Astronomy, University of California, Berkeley, CA 94720, USA}
\affil{Lawrence Berkeley National Laboratory, 1 Cyclotron Road, MS 50B-4206, Berkeley, CA 94720, USA}

\author[0000-0002-8532-9395]{Frank J. Masci}
\affil{IPAC, California Institute of Technology, 1200 E. California Blvd., Pasadena, CA 91125, USA}

\author[0000-0002-0387-370X]{Reed Riddle}
\affil{Caltech Optical Observatories, California Institute of Technology, Pasadena, CA 91125, USA}

\author[0000-0001-7062-9726]{Roger Smith}
\affil{Caltech Optical Observatories, California Institute of Technology, Pasadena, CA 91125, USA}

\begin{abstract}
The classification of variable objects provides insight into a wide variety of astrophysics ranging from stellar interiors to galactic nuclei. The Zwicky Transient Facility (ZTF) provides time-series observations that record the variability of more than a billion sources. The scale of these data necessitates automated approaches to make a thorough analysis. Building on previous work, this paper reports the results of the ZTF Source Classification Project (\scope), which trains neural network and XGBoost machine-learning (ML) algorithms to perform dichotomous classification of variable ZTF sources using a manually constructed training set containing 170,632 light curves. We find that several classifiers achieve high precision and recall scores, suggesting the reliability of their predictions for 209,991,147 light curves across 77 ZTF fields. We also identify the most important features for XGB classification and compare the performance of the two ML algorithms, finding a pattern of higher precision among XGB classifiers. The resulting classification catalog is available to the public, and the software developed for \scope\ is open-source and adaptable to future time-domain surveys.
\end{abstract}



\section{Introduction}
\label{sec:introduction}

Studying the variability of astronomical objects offers valuable insight into several open astrophysical questions, including the nature of stellar interiors \citep[e.g.][]{goupil2013_book_stellar_interiors}, dynamical interactions \citep[e.g.][]{borkovits_review_stellar_dynamics}, magnetic fields \citep[e.g.][]{fabbian2017_review_stellar_magnetic_activity}, cosmic distances \citep[e.g.][]{fukugita1993_review_cosmic_distances}, and galactic nuclei \citep[e.g.][]{ulrich1997_review_agn}. Surveys such as ASAS \citep{pojmanski2002_asas}, NSVS \citep{wozniak2004_nsvs, hoffman2009_nsvs}, PTF \citep{law2009_ptf}, the Catalina Surveys \citep{drake2014_catalina, drake2017_catalina}, and ASAS-SN \citep{kochanek2017_asassn, jayasinghe2018_asassn} have provided photometric time series data for millions of sources, facilitating the classification of variables and subsequent analyses. Over time, surveys have been further optimized for these purposes, covering more sky, reaching fainter limiting magnitudes, and observing at a faster cadence than their predecessors.

One such survey, the Zwicky Transient Facility \citep[ZTF;][]{bellm2019_ztf, graham2019_ztf, masci2019_ztf, dekany2020_ztf}, covers the full observable sky from Palomar Mountain every two nights, providing time-series data for billions of sources. The survey's latest data release (DR20, spanning 2018 March -- 2023 October) includes 53.5 million single-exposure images yielding 4.75 billion light curves spanning $g$, $r$, and $i$ bands. These data offer valuable insights into time-domain astronomy but also present unique challenges associated with their vast scale. While some manual studies have been performed for specific kinds of objects \citep[e.g.][]{vanroestel2021_amcvn}, a fully manual analysis of a survey this size demands unrealistic human resources.

Any large-scale analysis therefore requires an automated approach \citep[e.g.][]{huijse2014_ml_review, sen2022_ml_review}. One such approach involves the use of machine-learning (ML) algorithms at a survey-level scale \citep{mahabal2019}. ML can be used to direct follow-up observations based on the latest survey data \citep[e.g.][]{sravan2023}. ML techniques have also been applied to large surveys to classify different kinds of sources, including microlensing events \citep[e.g.][]{godines2019}, transients \citep[e.g.][]{stachie2020, gomez2023, rehemtulla2023_aas}, and variables \citep[e.g.][]{richards2011, garciajara2022, mistry2022}.

The major variable source classification effort within ZTF is the Source Classification Project (\scope). Beginning with a set of sources built from existing catalogs and light-curve classifications from human experts, \scope\ maps light-curve data to standardized features, trains many binary classifiers\footnote{Hereafter called ``dichotomous classifiers'' to avoid confusion with astrophysical binaries} using two supervised ML algorithms, and runs inference on unclassified sources with the goal of producing a variable source catalog for ZTF. These efforts have produced intermediate results reported in previous papers: \citet{scope_paper1} established the framework to train ML algorithms and run inference on an earlier ZTF data release. \citet{scope_paper2} set the light-curve variability metrics and period-finding algorithms that comprise the features input to ML algorithms. 

In this paper, we build on the foundation of these previous works to construct a publicly available ZTF variable source catalog containing ML classifications. Section \ref{sec:feature_generation} outlines the classification workflow and enumerates the standardized features generated from ZTF light curves. Section \ref{sec:ml_algorithm_training} describes the classification taxonomy and the two ML algorithms that label ZTF sources. Section \ref{sec:results} shares classifier performance and describes the resulting catalog of ZTF variables. We discuss these results in Section \ref{sec:discussion} before concluding in Section \ref{sec:conclusion}.

\section{Feature Generation}
\label{sec:feature_generation}

\begin{figure*}
    \centering
    \includegraphics[scale=0.25]{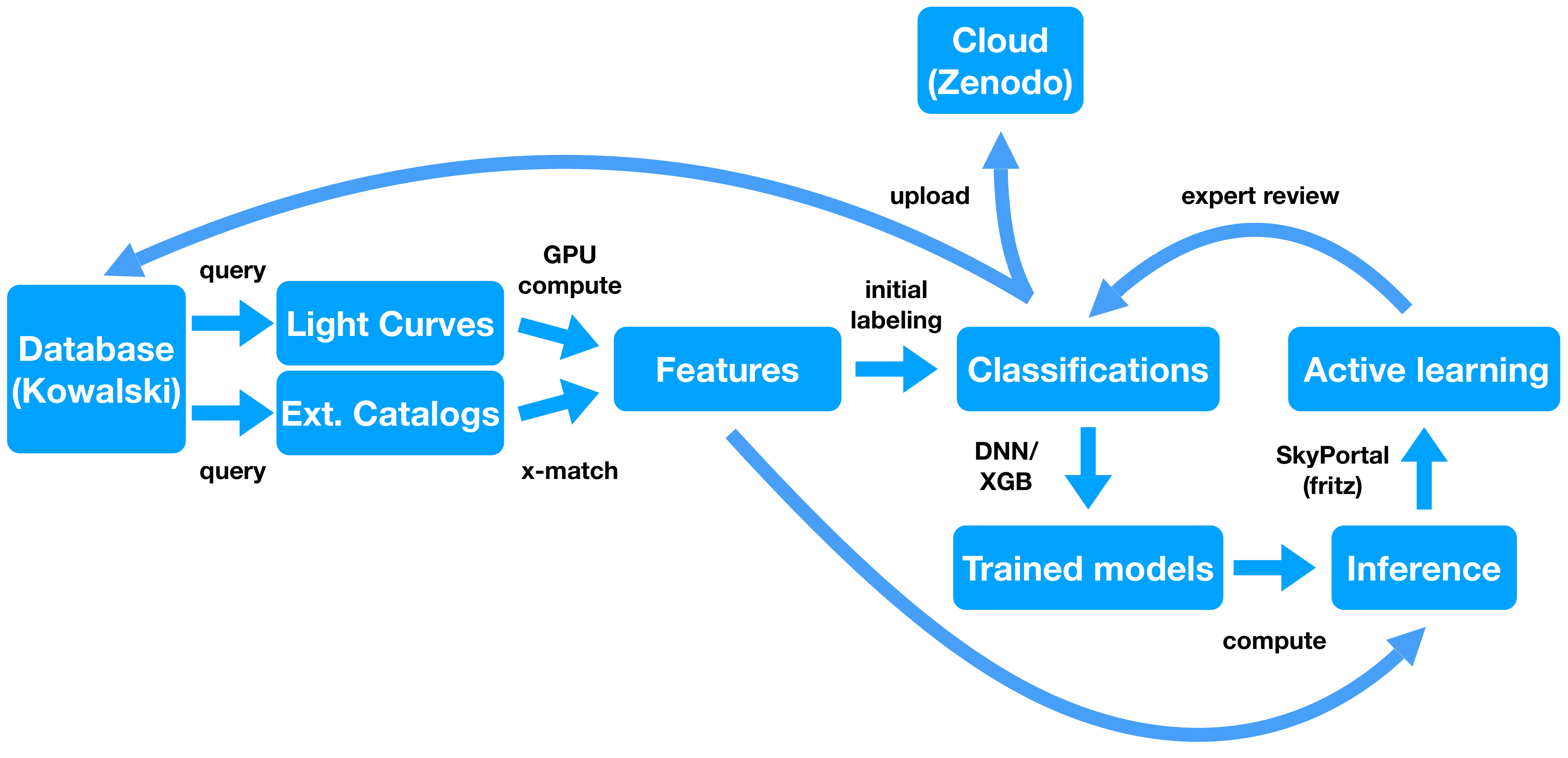}
    \caption{Workflow for SCoPe showing initial database queries, feature generation, and initial labeling followed by cycles of training, inference, and active learning. The training set and inference results are available publicly on Zenodo.}
    \label{fig:scope-workflow}
\end{figure*}

Figure \ref{fig:scope-workflow} depicts the \scope\ workflow that produces publicly available classifications starting with ZTF light curves on a \texttt{kowalski}\footnote{\url{https://github.com/skyportal/kowalski}} mongoDB database \citep{duev2019}. The \texttt{scope-ml} code underlying this workflow is available open-source on GitHub\footnote{\url{https://github.com/ZwickyTransientFacility/scope-ml}} and is published on PyPI\footnote{\url{https://pypi.org/project/scope-ml}}. Time-series data often have characteristics that make them unsuitable to be directly input to ML algorithms. These characteristics include multiple sampling rates, small gaps in the data due to changes in nightly observing conditions, and larger gaps due to a source's varying observability from Earth throughout the year (see Section 2.1 of \paperone). These light-curve qualities do not inform the astrophysical nature of sources but may still be learned by an ML classifier. To minimize the contribution of the above characteristics to source classification, we map each ZTF light curve to a set of standardized features to be input to our ML algorithms. This section describes in further detail these features and the process of their generation (covering all parts of the workflow in Figure \ref{fig:scope-workflow} leading to and including ``Features'').

\subsection{Selecting ZTF Sources}

We worked with ZTF DR16, which contains observations between 2018 March and 2023 January. The feature generation process begins with a query\footnote{Using \texttt{penquins} (\url{https://github.com/dmitryduev/penquins})} of either user-specified ZTF IDs or all ZTF sources within a single quadrant. This granulates the process into numerous parallelizable portions (four quadrants per CCD, 16 CCDs per field). We utilized the SDSC Expanse cluster to run many instances of our script in parallel. 

\subsection{Identifying Close, Bright Sources with Gaia}
Prior to computing features, we removed sources whose light curves may be influenced by nearby bright stars by querying Gaia EDR3 \citep{gaia2016, gaia2020_edr3}. We searched within a 300'' radius around each ZTF source, corresponding to the maximum separation that produces a light-curve artifact for sources with Tycho $B$ magnitudes of $B < 13$. Applying an empirical formula, we flagged ZTF sources for exclusion if we found neighboring stars bright enough to influence the source's light curve. We then queried light-curve data for the remaining sources.

\subsection{Light-curve Features and External Catalog Data}
We dropped all points from these light curves containing nonzero ZTF \texttt{catflags} (suggesting suboptimal data quality)\footnote{\url{https://irsa.ipac.caltech.edu/data/ZTF/docs/releases/ztf_release_notes_latest}}, and we subsequently enforced a 50-epoch minimum. For light curves meeting this requirement, we began by generating the basic statistics summarized in Table \ref{tab:phenom_features}. We also mapped each light curve to two-dimensional histograms showing the change in magnitude and time between each pair of points (\texttt{dmdt}, see Section 2.2.3 of \paperone\ and \citet{mahabal2017_dmdt} for more details). We supplemented these features with 2'' cross-match queries to ZTF Alerts and external catalogs. Additional features from these queries include the number and mean BRAAI of ZTF Alerts for each source \citep{duev2019} and a combination of magnitudes, errors, and parallax values from AllWISE \citep{wright2010_wise, cutri2021_allwise}, Pan-STARRS1 \citep[PS1;][]{kaiser2002_panstarrs, chambers2016_panstarrs1}, and Gaia EDR3 (Table \ref{tab:ontol_features}). 

\subsection{Period Finding and Fourier Features}
\label{subsec:period_finding}
We continued by running three period-finding algorithms on each light curve. These algorithms use GPU-accelerated implementations\footnote{\url{https://github.com/ZwickyTransientFacility/periodfind}} of Lomb--Scargle \citep[LS,][]{lomb1976, scargle1982}, conditional entropy \citep[CE;][]{graham2013_conditionalentropy}, and analysis of variance \citep[AOV;][]{schwarzenberg-czerny1998} methods to determine periods \textbf{and associated significance values}. The three algorithms ran on a grid of periods between 30 minutes and half the longest time baseline from among each batch of 1000 light curves. We excluded period ranges associated with common aliases in ground-based data, including several multiples of 1 day and 1 yr, along with a $\sim$ 30-day period for the Moon's orbit (see \papertwo; \citealt{kramer2023}). For each algorithm, the period with the highest significance value is reported as that algorithm's associated \texttt{period} feature.

We also applied a fourth algorithmic approach (ELS\_ECE\_EAOV) that nested the previous three algorithms. For each light curve, we used the full AOV results to determine the significance values associated with periods having the highest 50 LS and CE significance values. We then selected the period with highest AOV significance from this subset of 100 values. Using the resulting periods from all four algorithms, we generated additional features from the parameters of Fourier series fits (from zeroth to fifth order) to each light curve (see Eq. 1 of \paperone). We chose the ELS\_ECE\_EAOV algorithm to source the single set of Fourier features and periods input to the ML algorithms and reported with  classification predictions, since ELS\_ECE\_EAOV combines the results of each individual algorithm.

\begin{table*}
\begin{center}
\caption{Definitions of Features Input to Phenomenological and Ontological Classifiers.}
\label{tab:phenom_features}
\begin{tabular}{lc}
\hline
\hline
Feature Name & Definition \\
\hline
  \hline

    ad & Anderson--Darling statistic \citep{stephens1974_ad} \\
    chi2red & Reduced $\chi^{2}$ after mean subtraction \\
    f1\_BIC & Bayesian information criterion of best-fitting series \citep[Fourier analysis,][]{schwarz1978_f1_bic} \\
    f1\_a & $a$ coefficient of best-fitting series (Fourier analysis) \\
    f1\_amp & Amplitude of best-fitting series (Fourier analysis) \\
    f1\_b & $b$ coefficient of best-fitting series (Fourier analysis) \\
    f1\_phi0 & Zero-phase of best-fitting series (Fourier analysis) \\
    f1\_power & Normalized $\chi^{2}$ of best-fitting series (Fourier analysis) \\
    f1\_relamp1 & Relative amplitude, first harmonic (Fourier analysis) \\
    f1\_relamp2 & Relative amplitude, second harmonic (Fourier analysis) \\
    f1\_relamp3 & Relative amplitude, third harmonic (Fourier analysis) \\
    f1\_relamp4 & Relative amplitude, fourth harmonic (Fourier analysis) \\
    f1\_relphi1	& Relative phase, first harmonic (Fourier analysis) \\
    f1\_relphi2 & Relative phase, second harmonic (Fourier analysis) \\
    f1\_relphi3 & Relative phase, third harmonic (Fourier analysis) \\
    f1\_relphi4 & Relative phase, fourth harmonic (Fourier analysis) \\
    i60r & Mag ratio between 20th, 80th percentiles \\
    i70r & Mag ratio between 15th, 85th percentiles \\
    i80r & Mag ratio between 10th, 90th percentiles \\
    i90r & Mag ratio between 5th, 95th percentiles \\
    inv\_vonneumannratio & Inverse of von Neumann ratio \citep{vonneumann1941, vonneumann1942} \\
    iqr & Mag ratio between 25th, 75th percentiles \\
    median & Median magnitude \\
    median\_abs\_dev & Median absolute deviation of magnitudes \\
    norm\_excess\_var & Normalized excess variance \citep{nandra1997_norm_excess_var} \\
    norm\_peak\_to\_peak\_amp & Normalized peak-to-peak amplitude \citep{sokolovsky2009_norm_peak_to_peak_amp} \\
    roms & Robust median statistic \citep{rose2007_roms} \\
    skew & Skew of magnitudes \\
    smallkurt & Kurtosis of magnitudes \\
    stetson\_j & Stetson $J$ coefficient \citep{stetson1996} \\
    stetson\_k & Stetson $K$ coefficient \citep{stetson1996} \\
    sw & Shapiro--Wilk statistic \citep{shapiro1965} \\
    welch\_i & Welch $I$ statistic \citep{welch1993_welch_i} \\
    wmean & Weighted mean of magnitudes \\
    wstd & Weighted standard deviation of magnitudes \\
    dmdt\footnote{Not input to XGB algorithm.} & Magnitude--time histograms \citep[26 x 26,][]{mahabal2017_dmdt} \\
  \hline
\hline
\end{tabular}
\end{center}
\end{table*}

\begin{table*}
\begin{center}
\caption{Definitions of Additional Features Input to Ontological Classifiers.}
\label{tab:ontol_features}
\begin{tabular}{lc}
\hline
\hline
Feature Name & Definition \\
\hline
  \hline
    mean\_ztf\_alert\_braai & Mean significance of ZTF alerts for this source \citep{duev2019} \\
    n\_ztf\_alerts & Number of ZTF alerts for this source \citep{duev2019} \\
    period & Period determined by subscripted algorithms (e.g. ELS\_ECE\_EAOV) \\
    significance & Significance of period \\
    AllWISE\_w1mpro & AllWISE $W1$ mag \\
    AllWISE\_w1sigmpro & AllWISE $W1$ mag error \\
    AllWISE\_w2mpro & AllWISE $W2$ mag \\
    AllWISE\_w2sigmpro & AllWISE $W2$ mag error \\
    AllWISE\_w3mpro & AllWISE $W3$ mag \\
    AllWISE\_w4mpro & AllWISE $W4$ mag \\
    Gaia\_EDR3\_\_parallax & Gaia parallax \\
    Gaia\_EDR3\_\_parallax\_error & Gaia parallax error \\
    Gaia\_EDR3\_\_phot\_bp\_mean\_mag & Gaia $BP$ mag \\
    Gaia\_EDR3\_\_phot\_bp\_rp\_excess\_factor & Gaia $BP-RP$ excess factor \\
    Gaia\_EDR3\_\_phot\_g\_mean\_mag & Gaia $G$ mag \\
    Gaia\_EDR3\_\_phot\_rp\_mean\_mag & Gaia $RP$ mag \\
    PS1\_DR1\_\_gMeanPSFMag & PS1 $g$ mag \\
    PS1\_DR1\_\_gMeanPSFMagErr & PS1 $g$ mag error \\
    PS1\_DR1\_\_rMeanPSFMag & PS1 $r$ mag \\
    PS1\_DR1\_\_rMeanPSFMagErr & PS1 $r$ mag error \\
    PS1\_DR1\_\_iMeanPSFMag & PS1 $i$ mag \\
    PS1\_DR1\_\_iMeanPSFMagErr & PS1 $i$ mag error \\
    PS1\_DR1\_\_zMeanPSFMag & PS1 $z$ mag \\
    PS1\_DR1\_\_zMeanPSFMagErr & PS1 $z$ mag error \\
    PS1\_DR1\_\_yMeanPSFMag & PS1 $y$ mag \\
    PS1\_DR1\_\_yMeanPSFMagErr & PS1 $y$ mag error \\
  \hline
\hline
\end{tabular}
\end{center}
\end{table*}

\section{ML Algorithm Training}
\label{sec:ml_algorithm_training}

\subsection{Ontological and Phenomenological Taxonomies}
The goal of \scope\ is to use ML algorithms to reliably classify each ZTF source with as much detail as possible. The attainable quality of classifications varies across the broad range of ZTF sources. Factors that can affect the detail of source classifications include the quantity and quality of the data, the similarity of the training set to the source in question, and the existence of new kinds of variable sources in the data. With this in mind, we adopt two taxonomies that contain the labels we use to classify ZTF sources.

The first taxonomy is ontological (Figure \ref{fig:ontol-hierarchy}) and contains specific kinds of astrophysical sources (see Table \ref{tab: classifier_performance} for the ontological labels, training set abbreviations, and definitions, ordered by low to high detail). This list aims to include as many kinds of objects as feasible for expert classification review (see Section \ref{subsubsec:active_learning}). When training ontological classifiers, we input the full set of features (Tables \ref{tab:phenom_features} and \ref{tab:ontol_features}) to the ML algorithms. 

In consideration of the value of having some information about a source (even if not a definitive ontological classification), we also employed a phenomenological taxonomy (Figure \ref{fig:phenom-hierarchy}) with labels that describe light-curve-based features. Classifications with (p) in their definition in Table \ref{tab: classifier_performance} denote the phenomenological labels with their training set abbreviations and definitions. Phenomenological classifiers trained on the phenomenological subset of features (Table \ref{tab:phenom_features}) to ensure that their classification results were only dependent on ZTF light curves.

\subsection{Dichotomous Classifiers}

\begin{figure*}
    \centering
    \includegraphics[scale=0.75]{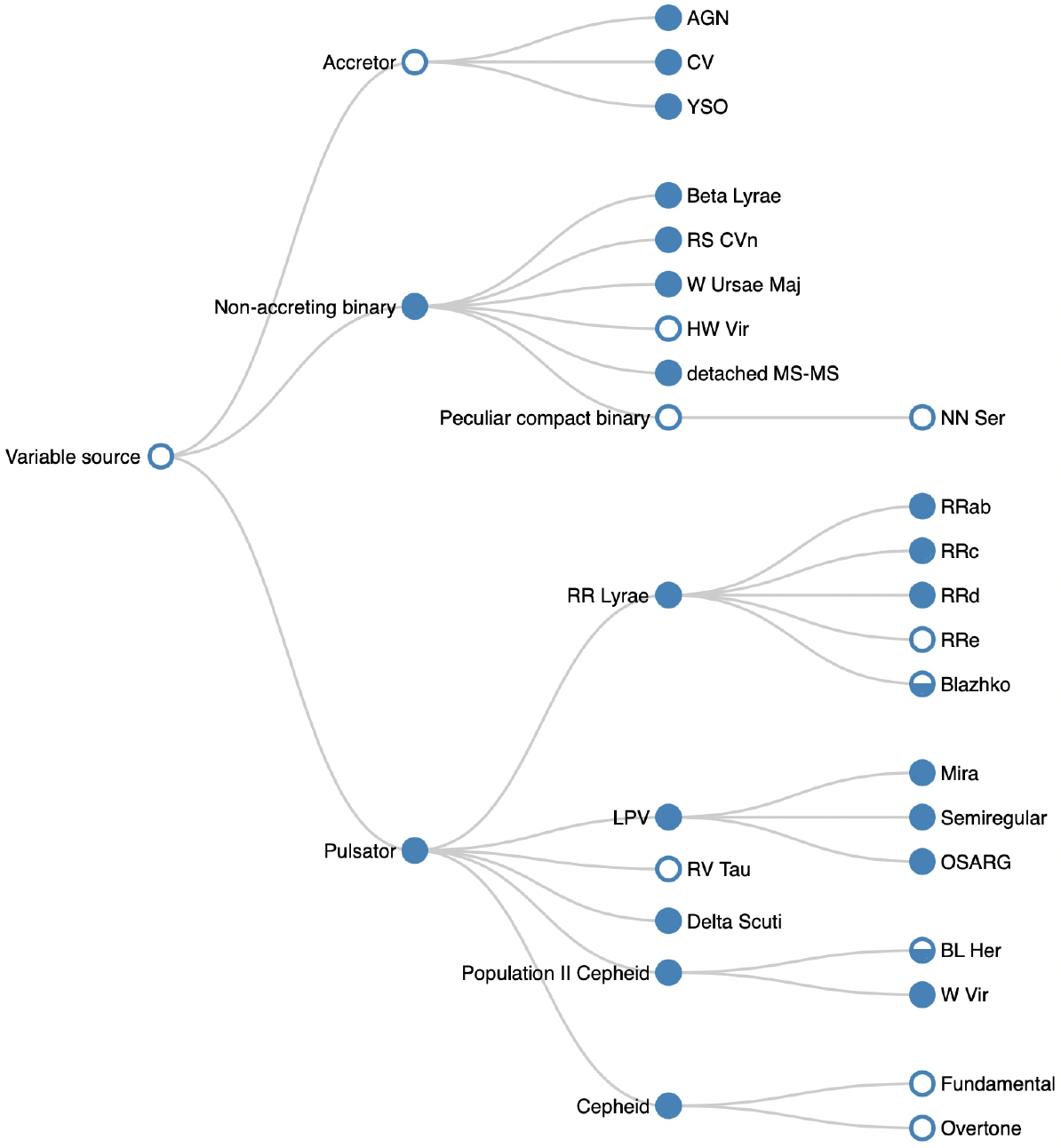}
    \caption{Hierarchy of ontological classes used in \scope. Not intended to be an exhaustive taxonomy, this collection of labels organizes the intrinsic classifications that compose the training set. Filled circles indicate labels for which we trained a classifier. Open circles show labels that were not used for training owing to their few ($<$ 50) positive examples or solely organizational nature (``Variable source,'' ``Accretor''). Half-filled circles identify labels with enough positive examples but lacking a successfully trained classifier (see Section \ref{subsec:training}).}
    \label{fig:ontol-hierarchy}
\end{figure*}

\begin{figure*}
    \centering
    \includegraphics[scale=0.725]{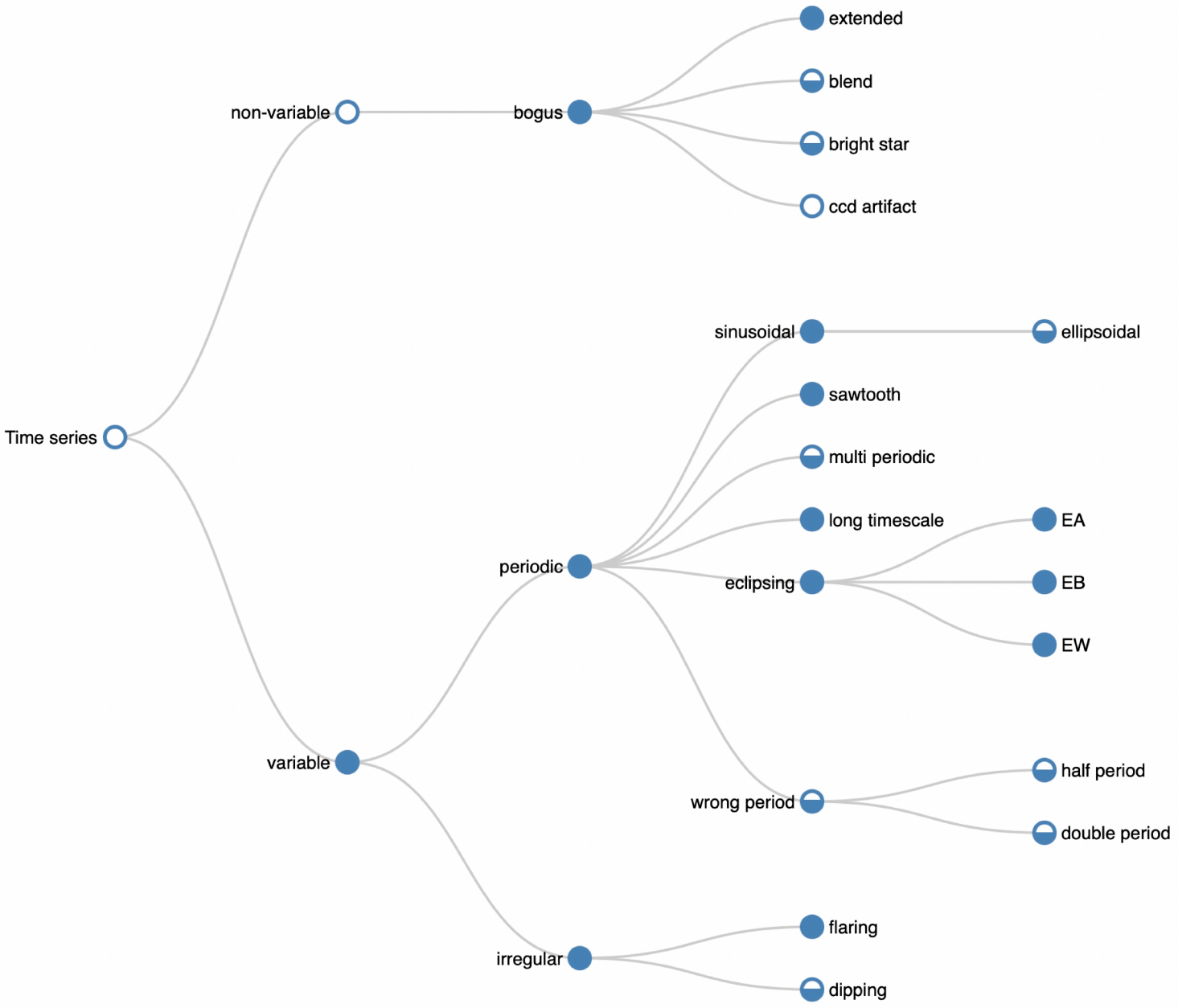}
    \caption{Hierarchy of phenomenological classes in  \scope, following the filled-circle pattern of Figure \ref{fig:ontol-hierarchy}. As with the ontological hierarchy, only non--top-level classes having 50 or more positive examples were used for training. Additionally, ``nonvariable'' is an organizational label, since the nonvariable probability is defined as  1 $-$ the ``variable'' probability.}
    \label{fig:phenom-hierarchy}
\end{figure*}

We trained independent dichotomous classifiers for labels in these taxonomies having more than 50 positive examples. The choice of dichotomous classifiers allows more than one label to be assigned to a source, often with varying levels of detail. This is important not only because of the practical challenges outlined above but also because some sources merit more than one classification (e.g. an eclipsing binary system containing a flaring star). The independence of dichotomous classifiers allows for future updates to the taxonomies without a revision of the current results from each existing classifier. Although dichotomous classifiers each only consider one label, we used the hierarchical structure of our taxonomies to assist in filling missing labels before training (see Figures \ref{fig:ontol-hierarchy} and \ref{fig:phenom-hierarchy} along with Section \ref{subsubsec:upstream_labels}).

\subsection{ML Algorithms}

We employed a convolutional/dense neural network \citep[hereafter DNN; e.g.][]{lecun2015_deeplearning} and XGBoost gradient-boosted decision trees \citep[XGB;][]{chen2016_xgboost} to perform classification (leading to ``Trained models'' in Figure \ref{fig:scope-workflow}). We applied a probability threshold on the input classification probabilities to determine whether to treat each source as a positive or negative example for training. A threshold that is too high could impede a classifier's ability to generalize, while too low a threshold could include too many false positives. We chose to set the threshold at 0.7 for all classifiers, thus treating moderate-to-high-confidence probabilities as positive examples during training.

Both algorithms initially performed regression to minimize a binary cross-entropy loss function, assigning a classification probability ranging between 0 and 1 for each source. We again used a probability threshold of 0.7 to map predicted classifications for training sources in order to quantify true and false positives/negatives. When a source was associated with multiple ZTF light curves, the same classification probabilities were assigned to each one. We then trained classifiers on the collection of light curves. We train classifiers on a mix of ZTF bands in an effort to make them robust to the systematics between light curves in each band.

\subsubsection{Deep Neural Network (DNN)}
Neural networks map input to output using connected layers of artificial ``neurons'' inspired by their biological counterparts; each neuron performs a linear transformation of its input followed by a nonlinear activation function. The output from one neuron becomes the input of the next, and network training occurs via back-propagation of the loss function gradient through each possible path in the network. This process optimizes the weight and bias values the network's neurons use for linear transformations.

The \scope\ DNN algorithm features two branches built using \texttt{tensorflow}\footnote{\url{https://github.com/tensorflow/tensorflow}} with the \texttt{keras} API\footnote{\url{https://www.tensorflow.org/guide/keras}}: the first is a series of fully connected dense layers interspersed with dropout layers. This branch receives all features for a given classifier except \texttt{dmdt}. Each dropout layer randomly sets a fraction of inputs to zero in the subsequent layer. This regularization process helps prevent overfitting, wherein the network only learns specific details of the training set at the expense of performance on unseen data.

\begin{figure*}
    \centering
    \includegraphics[scale=0.125]{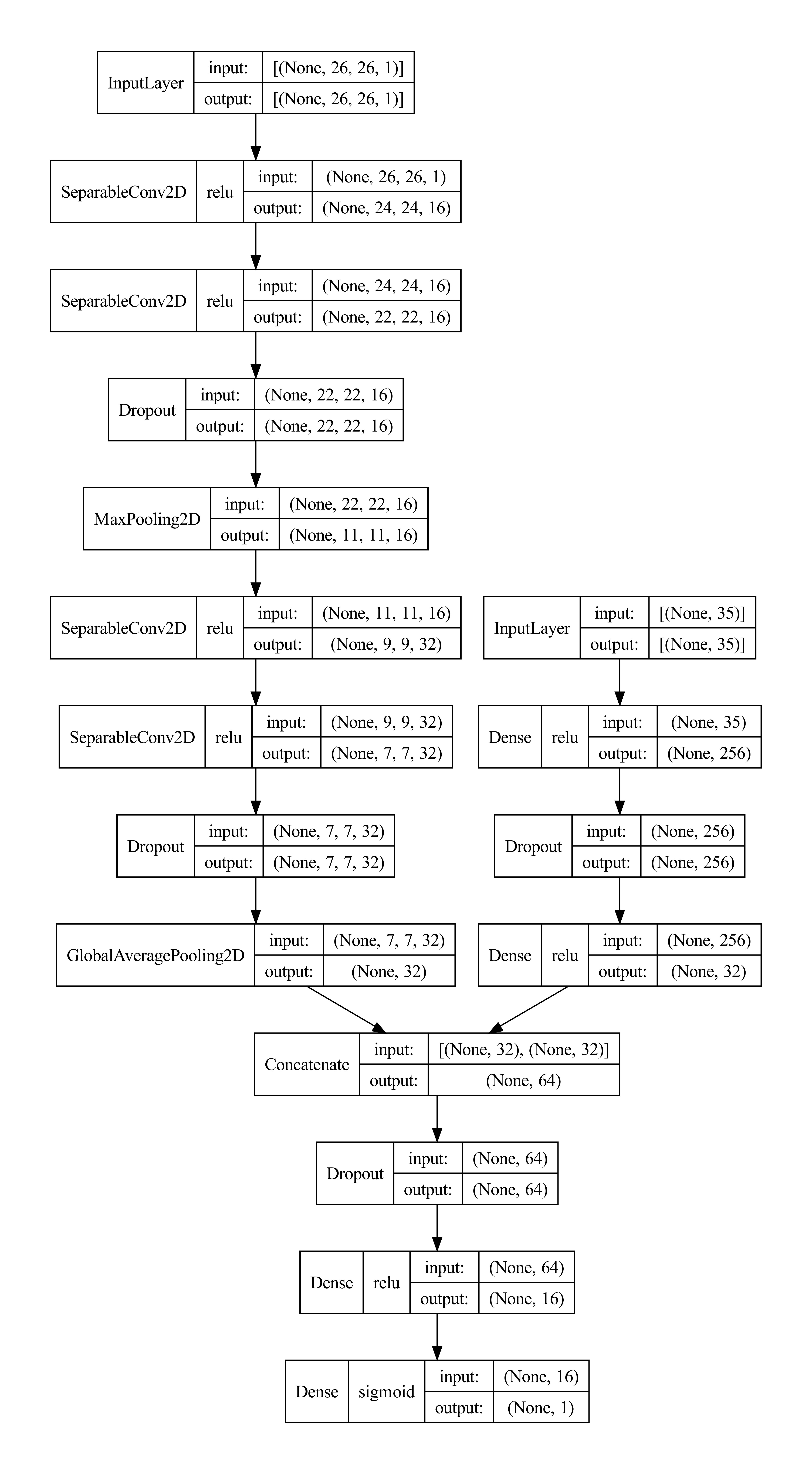}
    \caption{DNN architecture graph showing \texttt{keras} layer names and activation functions along with the shapes of their input and output. ``None'' dimensions indicate the networks's ability to work with any number of sources. The network combines a convolutional branch (taking \texttt{dmdt} as input) and fully connected layers (taking all other features) to yield classification probabilities. Dropout layers introduce regularization, helping to reduce overfitting.}
    \label{fig:dnn_architecture}
\end{figure*}

The other DNN branch convolves a kernel function with the 2D \texttt{dmdt} histograms. This process, commonly used for image analysis tasks, generates a set of ``features'' for \texttt{dmdt}. Dropout and pooling layers provide regularization for this branch. The outputs of both branches are concatenated and fed through one more set of dropout and dense layers before passing through a sigmoid activation function to provide continuous outputs between 0 and 1. These outputs correspond to one classification probability per source passed into the network. Figure \ref{fig:dnn_architecture} shows a graph of DNN branches and layers.

\subsubsection{Gradient-boosted Decision Trees (XGB)}
\label{subsubsec:xgboost}
The XGB algorithm\footnote{\url{https://github.com/dmlc/xgboost}} takes a different approach to classification based on a collection of decision trees. Each tree makes splits in feature space in order to optimally separate positive and negative training examples. Instead of aggregating these results (a ``bagging'' approach that introduces regularization), XGB computes the gradient of the loss function with respect to the trees' predictions and uses this information to inform another round of training (a ``boosting'' approach). Boosting offers high performance while increasing the chance of overfitting.

The splits in the XGB tree-based approach facilitate a straightforward interpretation of the connection between input features and resulting classifications. As a result, the importance of each input feature in determining a classification is provided by each XGB classifier. This contrasts with DNN classifiers, whose hidden layer transformations preclude the same kind of interpretation without significant added computational expense. Phenomenological and ontological XGB classifiers were given the same feature sets as DNN classifiers except for the exclusion of the \texttt{dmdt} histograms. 

\subsection{Preparing the Training Set}

\subsubsection{Imputing Missing Features}
The above algorithms cannot process missing features, so we adopted a heterogeneous feature imputation strategy. This strategy arose from a feature-by-feature consideration of how to appropriately fill the different kinds of missing quantities in our data. We performed no imputation for any phenomenological features, instead excluding the single light curve in the training set that was missing any of these features. We imputed zero for \texttt{n\_ztf\_alerts} and \texttt{mean\_ztf\_alert\_braai}. We imputed the median for uncertainties in survey magnitude and Gaia parallax. Finally, we used K-nearest neighbor regression to impute missing magnitudes and parallax values. 

\subsubsection{Assigning Upstream Labels}
\label{subsubsec:upstream_labels}
It is possible for the labeling process to produce an incomplete list of classifications for a source. For example, a source may be confidently labeled as \texttt{periodic} without having been labeled as \texttt{variable}. This will provide an incorrect input to the \texttt{variable} classifier, since this example source that displays periodic variability will be treated as nonvariable during training. To address this issue, we used the hierarchy of labels in each taxonomy (Figures \ref{fig:ontol-hierarchy} and \ref{fig:phenom-hierarchy}) to enforce that any labels upstream of a manual classification must have at least the same probability as the manual classification.

\subsubsection{Active Learning}
\label{subsubsec:active_learning}
We ran several rounds of algorithm training over the course of the project in an effort to increase the quantity and quality of our training set. After each round, we ran inference on unclassified sources using the trained classifiers (``Inference'' in Figure \ref{fig:scope-workflow}). The predictions for each light curve were aggregated using the mean classification probabilities of all light curves sharing the same Gaia, AllWISE, or PS1 survey ID. This aggregation produced one set of predictions per ZTF source.

We selected a subset of sources having at least one high-confidence  probability ($> 0.9$) among the labels in our taxonomies. We used a hosted instance of the \texttt{SkyPortal}\footnote{\url{https://github.com/skyportal/skyportal}} data platform \citep[\texttt{fritz};][]{vanderwalt2019_skyportal, coughlin2023_skyportal} to visualize the light curves of this subset of sources along with any associated ML classifications having probability $> 0.7$. Including these moderate-confidence classifications allowed an evaluation of the classifier's decision-making threshold in addition to its highest-confidence results.

\begin{figure*}
    \centering
    \includegraphics[scale=0.365]{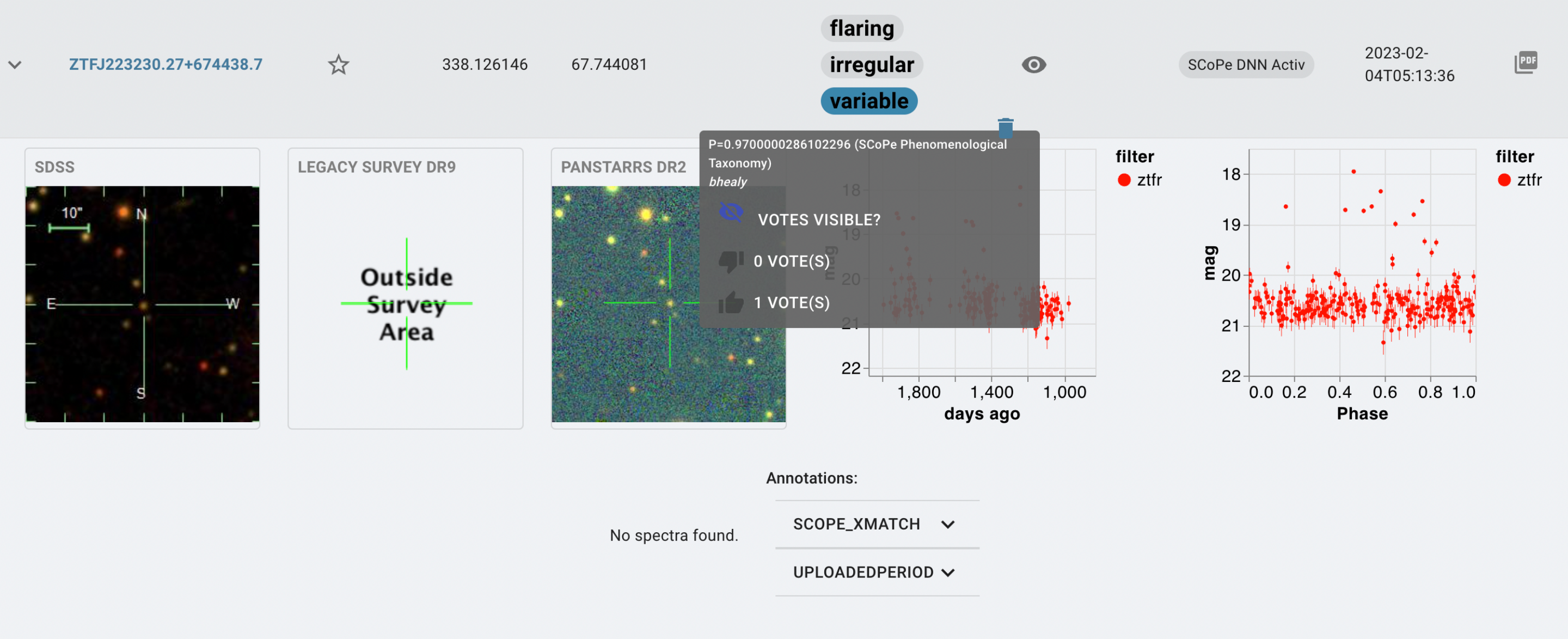}
    \includegraphics[scale=0.365]{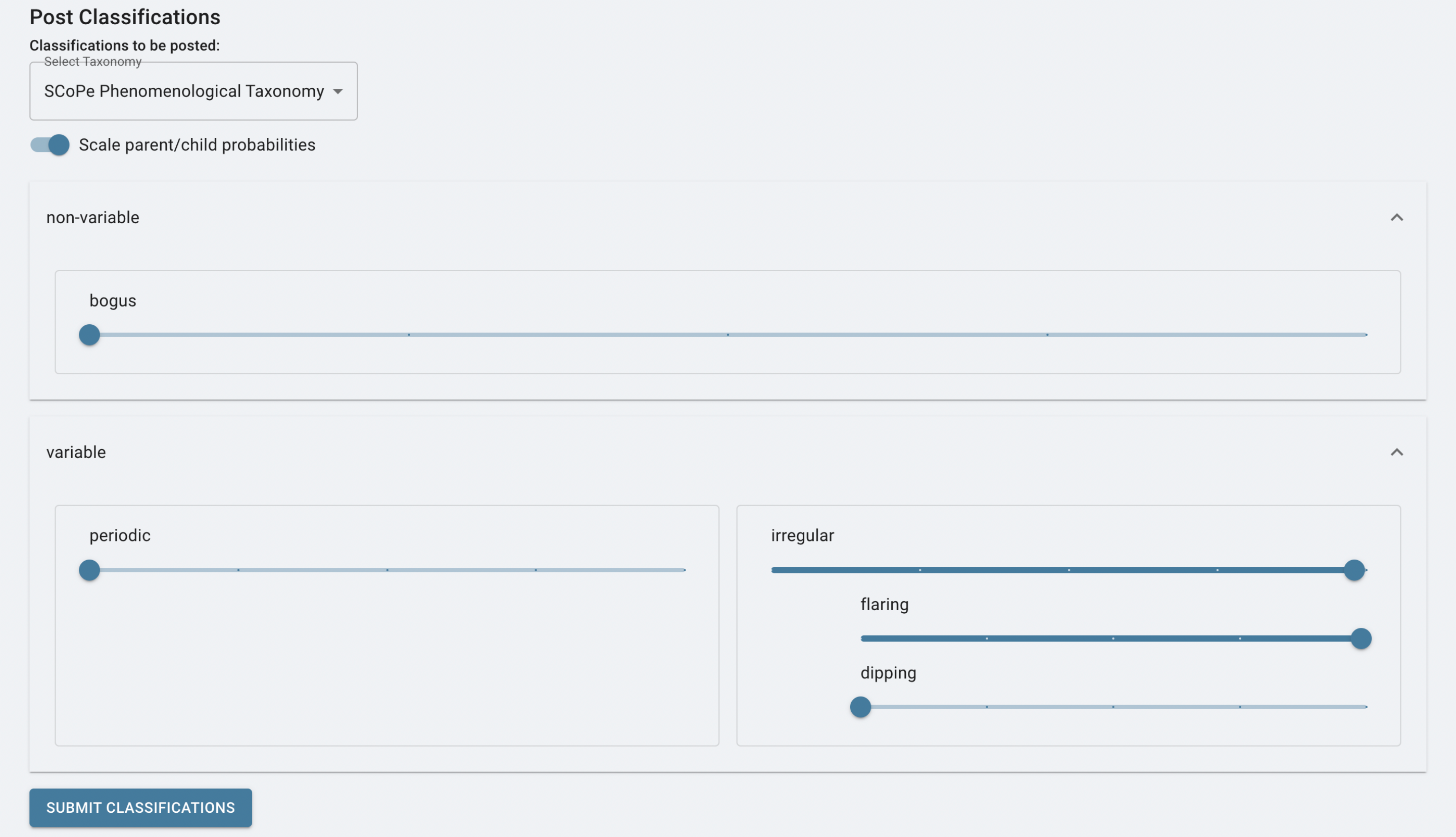}
    \caption{\texttt{fritz} interface used to vote, add, and remove classifications from sources. At top are the source's name, coordinates, and current classifications. Voting options become visible when mousing over a classification. Beneath are cutouts from the Sloan Digital Sky Survey, Legacy Survey, and Pan-STARRS, along with time-series and phase-folded photometry. Below is the slider interface used to add new classifications from a selected taxonomy.}
    \label{fig:fritz-labeling}
\end{figure*}

These sources were subjected to a round of human review to evaluate the classifier predictions and revise them as appropriate (``Active learning'' in Figure \ref{fig:scope-workflow}). Reviewers could vote ``up'' (+1) or ``down'' (-1) on each classification, leave no vote (0), and add any labels thought to be missing. Figure \ref{fig:fritz-labeling} shows an example of the \texttt{fritz} interface for voting and labeling sources.

Once this process was complete, we identified all sources reviewed by at least one human and summed the numerical votes to determine which classifications to keep. If the vote sum for a classification was negative, we removed it from the source. Otherwise, we added the source and its remaining classifications to the existing training set. In this way, we could increase the number of manually labeled examples provided for classifier training without requiring a random search of current predictions.

The training set initially began as a compilation of existing source catalogs and a ``seed'' set of manual classifications (``initial labeling'' in Figure \ref{fig:scope-workflow}). The cyclical process of active learning grew the training set to the point of containing manually reviewed labels for 85,136 sources along with features generated from 170,632 associated light curves, on which ran a final round of classifier training for all classes having 50 or more positive examples. The training set is available to the public electronically on Zenodo\footnote{\texttt{training\_set.parquet} at \url{https://zenodo.org/doi/10.5281/zenodo.8410825}}.

\subsection{Hyperparameter Optimization and Training}
We shuffled and divided the learning set into three partitions, reserving 81\% of rows for training, 9\% for model validation, and 10\% to obtain the test scores we report in this paper. We optimized DNN and XGB hyperparameters via two unique tuning processes on a subset of 9\% of the training data reserved for model validation.

For DNN, we used Weights and Biases Sweeps\footnote{\url{https://docs.wandb.ai/guides/sweeps}} to investigate changes in model performance for different parameter combinations. Specifically, we optimized the \texttt{amsgrad}, \texttt{epsilon} and \texttt{lr} hyperparameters. Training ran for 200 epochs. The optimal DNN hyperparameters for each classifier are saved in the default configuration file in the \scope\ code repository.

For XGB, we performed an initial piecemeal grid search separately optimizing up to two hyperparameters at a time (\texttt{max\_depth} and \texttt{min\_child\_weight}, \texttt{subsample} and \texttt{colsample\_bytree}, and \texttt{eta}). We followed these optimizations with one additional round each for the former two pairs of parameters using a more granular grid. XGB training then ran for up to 999 epochs unless there was no improvement in the area under the ROC curve for 10 consecutive rounds. XGB hyperparameter tuning can be reproduced by running the training code from the \scope\ repository using the training set on Zenodo. The results of training and inference are presented in the next section.

\section{Classification Results}
\label{sec:results}

\subsection{Training}
\label{subsec:training}
We report dichotomous classifier training results using precision and recall scores. The precision quantifies the fraction of sources labeled by the classifier as positive as compared to the number of true-positive examples in the test set (e.g. purity). The recall is the fraction of positive examples correctly labeled as positive by the classifier (e.g. completeness). Table \ref{tab: classifier_performance} shares the precision and recall scores for each DNN and XGB classifier.

Recall scores are zero in cases where an algorithm did not correctly classify any of the positive training examples. In these cases, the precision can be either undefined (no light curves classified as positive) or zero (only false-positive classifications). For some classes, both the DNN and XGB algorithms failed to achieve nonzero recall. These classes are listed at the bottom of Table \ref{tab: classifier_performance} and identified by half-filled circles in Figures \ref{fig:ontol-hierarchy} and \ref{fig:phenom-hierarchy}.

\begin{figure*}
    \centering
    \includegraphics[scale=0.45]{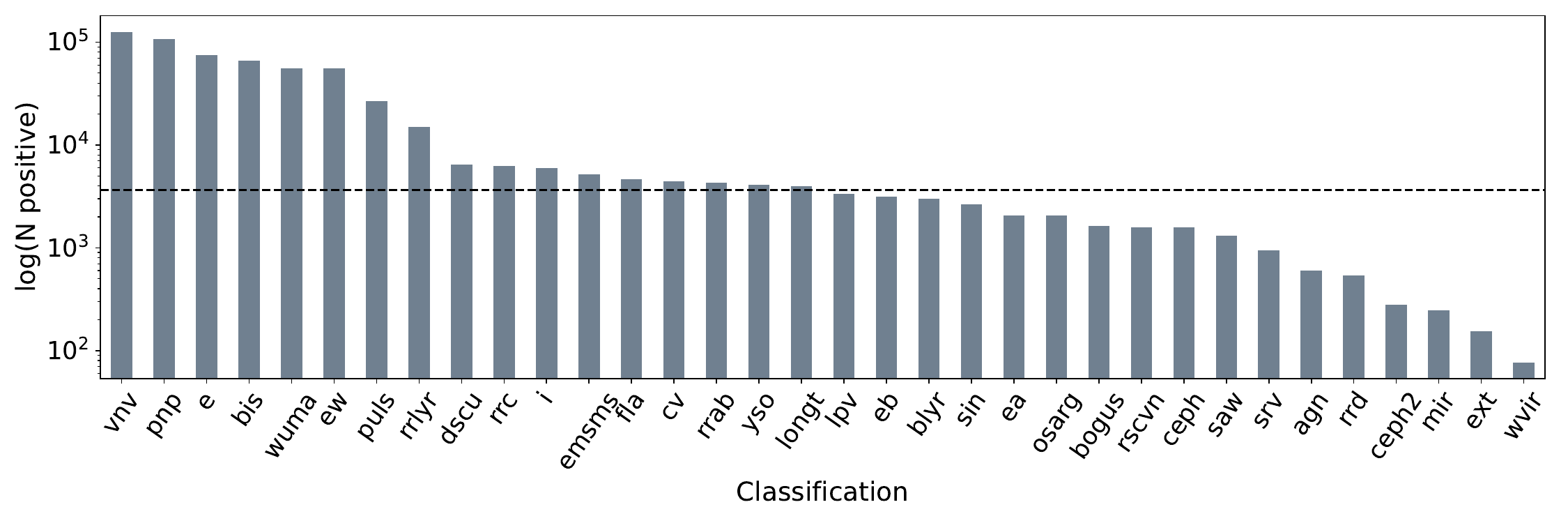}
    \caption{Number of positive examples for each classification in the learning set. The dashed line shows the median number of positive examples among all classes.}
    \label{fig:positive_example_counts}
\end{figure*}

\begin{figure*}
    \centering
    \includegraphics[scale=0.45]{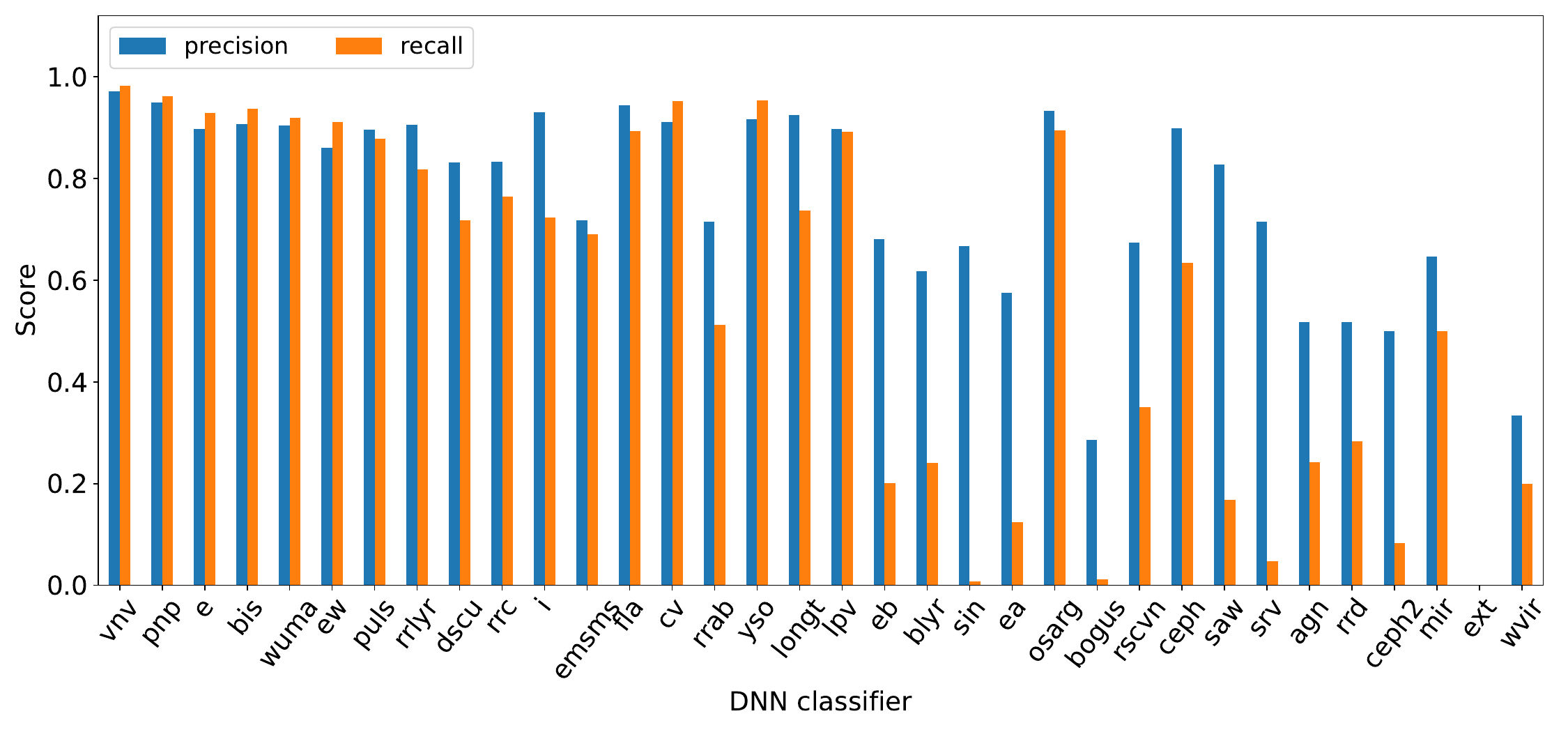}
    \includegraphics[scale=0.45]{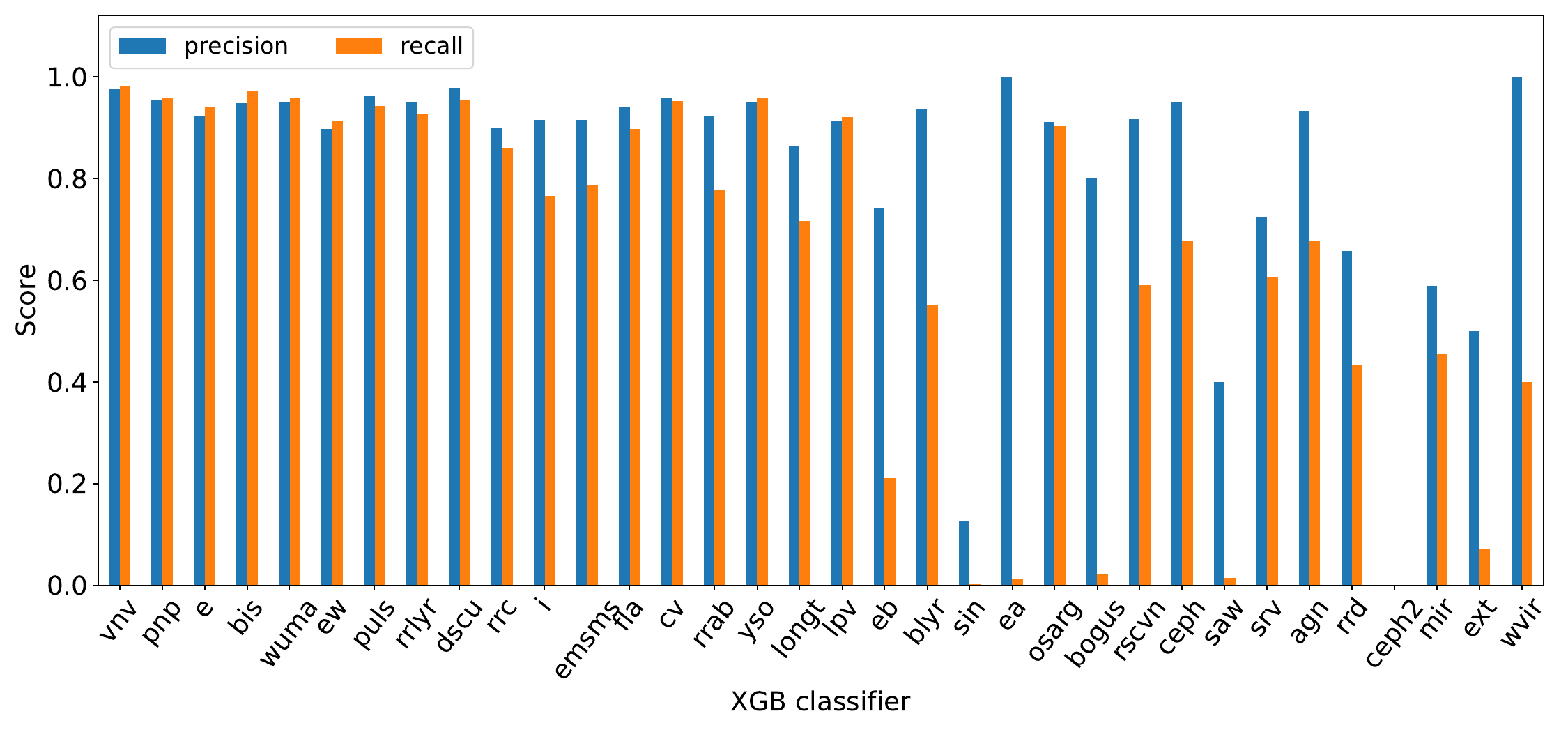}
    \caption{Test precision and recall stats for DNN (top) and XGB (bottom) classifiers. The number of positive examples in the training set decreases from left to right.}
    \label{fig:dnn_xgb_precision_recall}
\end{figure*}

\begin{figure*}
    \centering
    \includegraphics[scale=0.6]{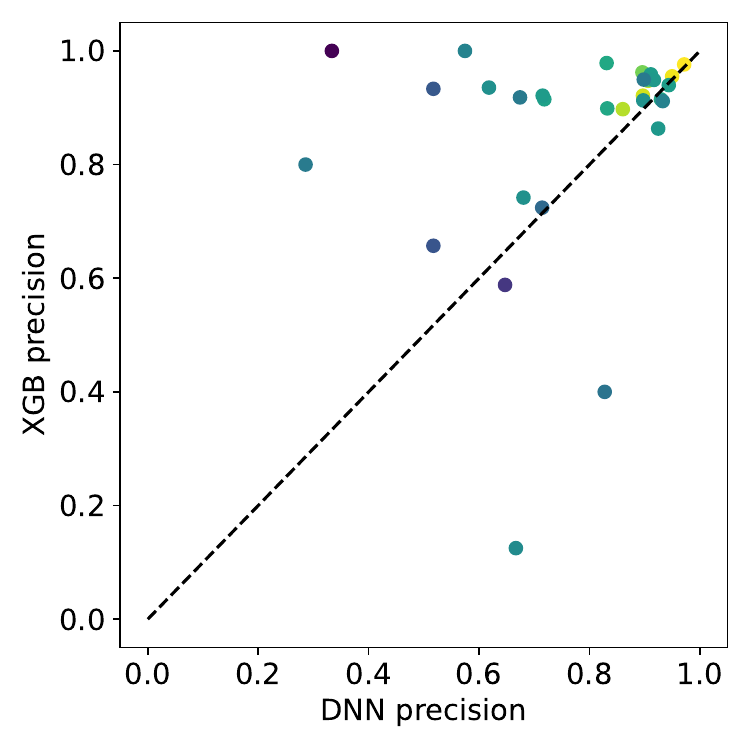}
    \includegraphics[scale=0.6]{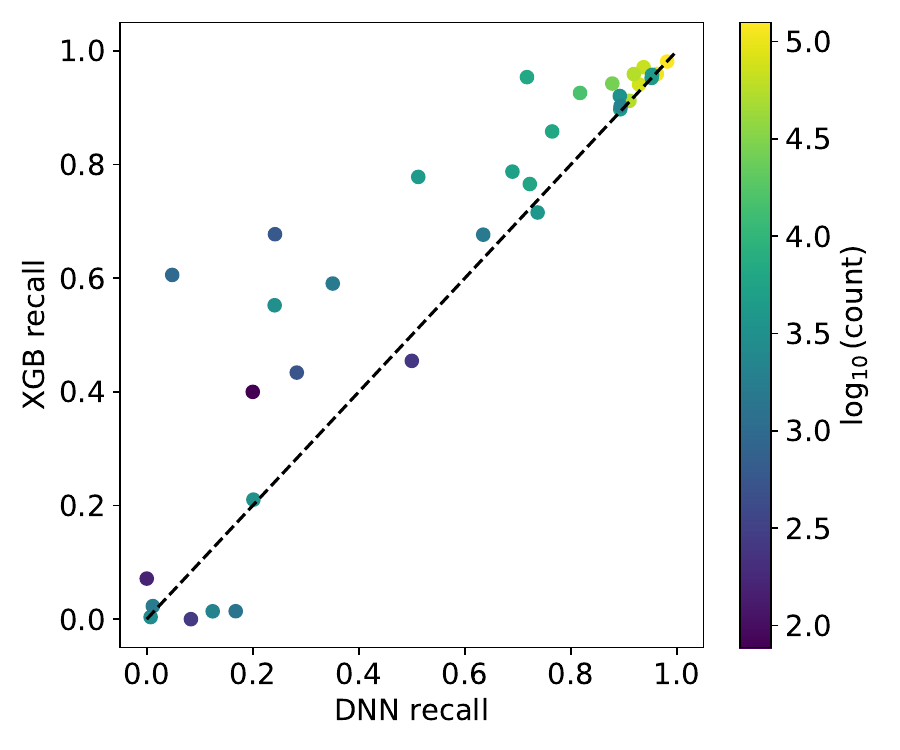}
    \caption{Scatter plots of XGB vs.\ DNN test precision (left) and recall (right) scores. Points are color-coded by the logarithm of the number of positive  examples in the learning set.}
    \label{fig:precision_recall_scatter}
\end{figure*}

\begin{figure*}
    \centering
    \includegraphics[scale=0.45]{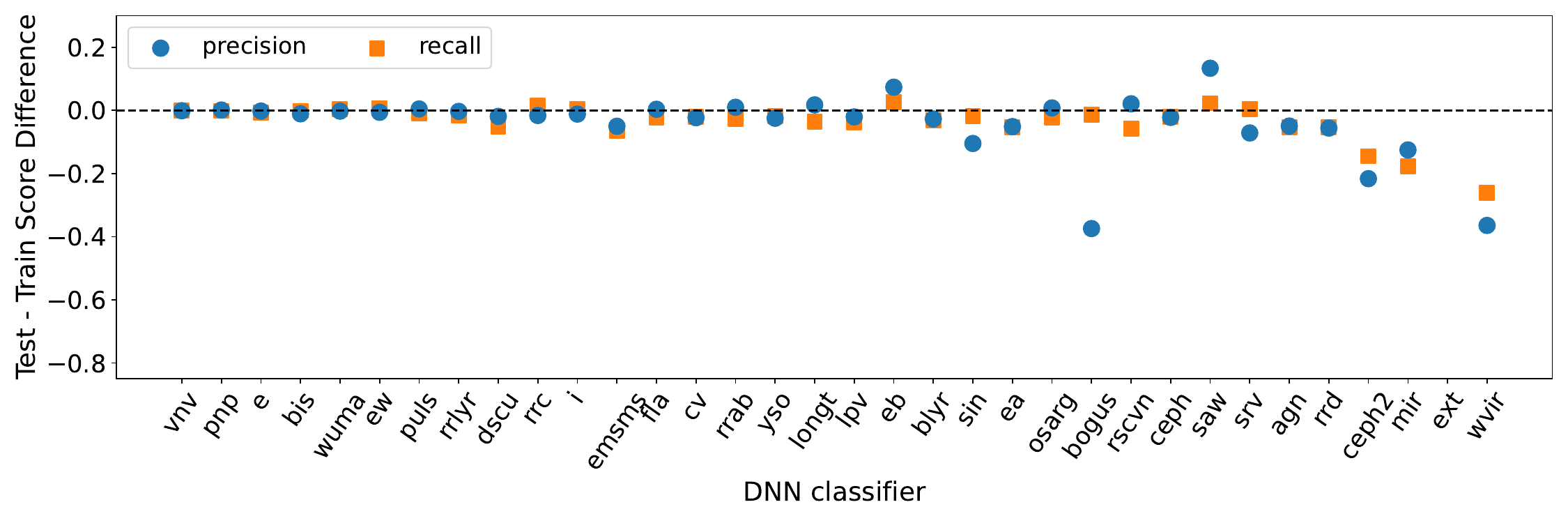}
    \includegraphics[scale=0.45]{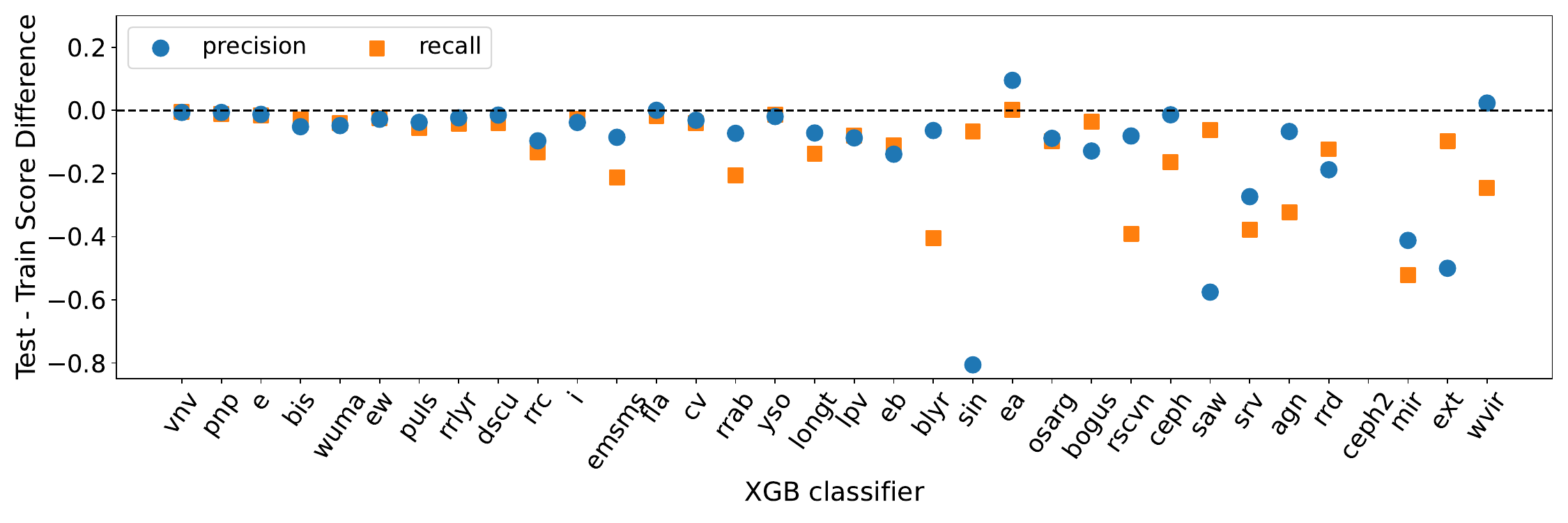}
    \caption{Difference between test and training set precision/recall scores for DNN (top) and XGB (bottom) classifiers. The difference in precision is indicated by blue circles, while the difference in recall is marked by orange squares. A significant negative difference between test and train scores indicates a higher chance of overfitting by the algorithm.}
    \label{fig:test_train_scores_diff}
\end{figure*}

\begin{figure*}
    \centering
    \includegraphics[scale=0.7]{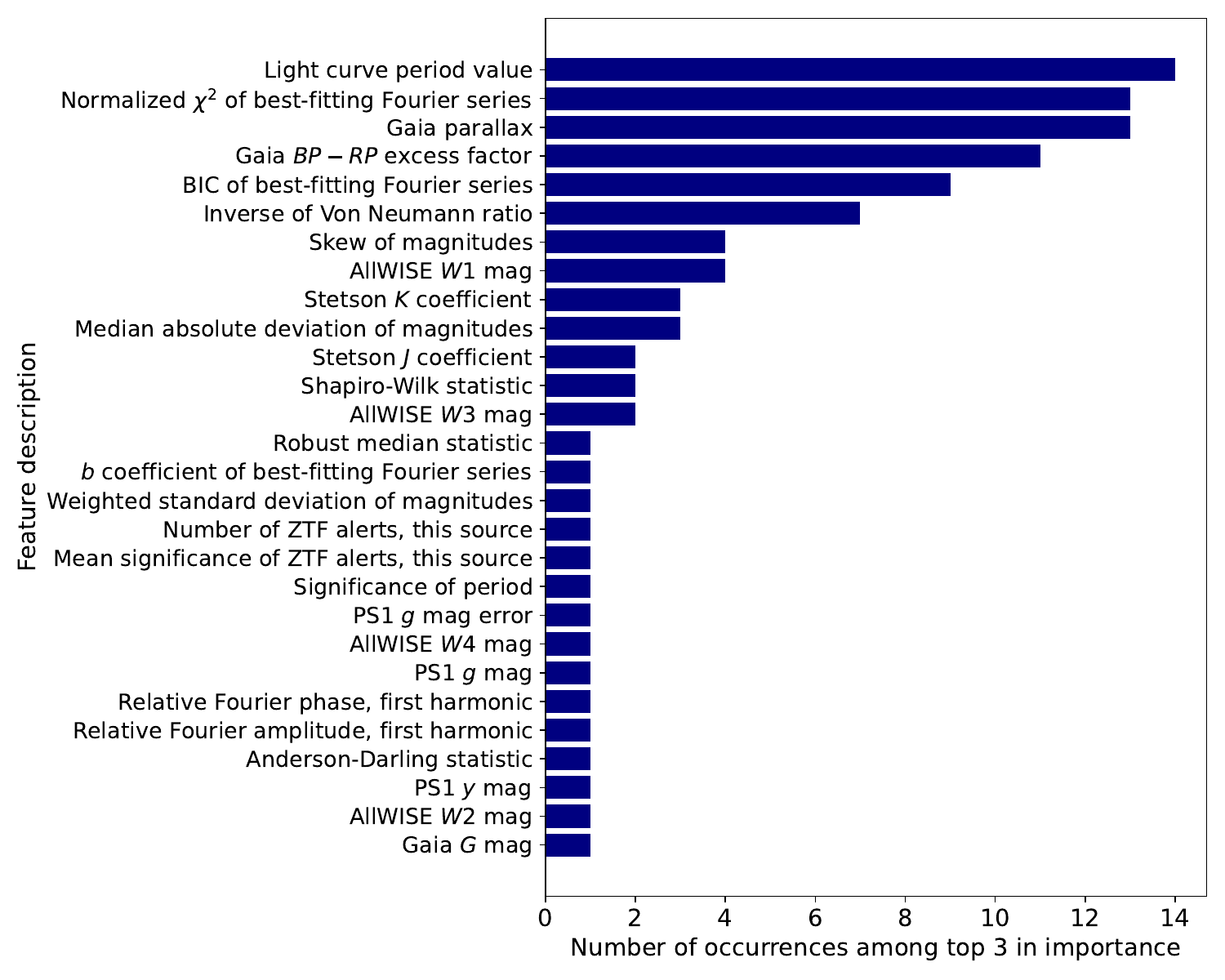}
    \caption{Number of occurrences of features among the top three in importance for each XGB classifier.}
    \label{fig:most_important_features_top3}
\end{figure*}

Figure \ref{fig:positive_example_counts} shows the number of positive examples for each class in the learning set and compares them to the median of all classes. The top and bottom panels of Figure \ref{fig:dnn_xgb_precision_recall} plot the precision/recall scores for DNN and XGB classifiers, respectively. Figure \ref{fig:precision_recall_scatter} shows a scatter-plot comparison of DNN and XGB precision/recall scores with color-coding for the number of positive training examples. Figure  \ref{fig:test_train_scores_diff} plots the difference between test and train scores to provide quantitative insight into overfitting. Figure \ref{fig:most_important_features_top3} shows a histogram counting the number occurrences of each feature among the top three in importance for each XGB classifier.

\begin{table*}
\begin{center}
\caption{Classification Abbreviations, Names, Definitions, Number of Positive Training Examples, Precision, and Recall.}
\label{tab: classifier_performance}
\begin{tabular}{lcccccc}
\hline
\hline
Classif. & Definition & No. Positive & DNN Precision & DNN Recall & XGB Precision & XGB Recall \\
\hline
  \hline
    vnv & variable (p) & 125,056 & 0.97 & 0.98 & 0.98 & 0.98 \\
    pnp & periodic (p) & 107,248 & 0.95 & 0.96 & 0.96 & 0.96 \\
    e & eclipsing (p) & 75,049 & 0.90 & 0.93 & 0.92 & 0.94 \\
    bis & nonaccreting binary system & 66,607 & 0.91 & 0.94 & 0.95 & 0.97 \\
    wuma & W UMa binary system & 55,253 & 0.90 & 0.92 & 0.95 & 0.96 \\
    ew & EW eclipsing (p) & 55,251 & 0.86 & 0.91 & 0.90 & 0.91 \\
    puls & pulsating star & 26,603 & 0.91 & 0.88 & 0.96 & 0.94 \\
    rrlyr & RR Lyr star & 14,917 & 0.90 & 0.82 & 0.95 & 0.93 \\
    dscu & Delta Scu star & 6475 & 0.83 & 0.72 & 0.98 & 0.95 \\
    rrc & RR Lyr c star & 6251 & 0.83 & 0.76 & 0.90 & 0.86 \\ 
    i & irregular variability (p) & 5970 & 0.93 & 0.72 & 0.91 & 0.77 \\
    emsms & eclipsing MS--MS binary & 5199 & 0.72 & 0.69 & 0.91 & 0.79 \\
    fla & flaring (p) & 4633 & 0.94 & 0.89 & 0.94 & 0.90 \\
    cv & cataclysmic variable & 4435 & 0.91 & 0.95 & 0.94 & 0.90 \\
    rrab & RR Lyr ab star & 4293 & 0.72 & 0.51 & 0.92 & 0.78 \\
    yso & young stellar object & 4108 & 0.92 & 0.95 & 0.95 & 0.96 \\
    longt & long-timescale variability (p) & 3961 & 0.92 & 0.74 & 0.86 & 0.72 \\
    lpv & long-period variable star & 3359 & 0.90 & 0.89 & 0.91 & 0.92 \\
    eb & EB eclipsing (p) & 3130 & 0.68 & 0.20 & 0.74 & 0.21 \\
    blyr & Beta Lyr binary & 3017 & 0.62 & 0.24 & 0.94 & 0.55 \\
    sin & sinusoidal (p) & 2645 & 0.67 & 0.01 & 0.13 & 0.004 \\
    ea & EA eclipsing (p) & 2073 & 0.57 & 0.12 & 1.0 & 0.01 \\
    osarg & OGLE small-amplitude red giant star\footnote{\citet{wray2004_osarg}} & 2061 & 0.93 & 0.89 & 0.91 & 0.90 \\
    bogus & bogus variability (p) & 1640 & 0.29 & 0.01 & 0.80 & 0.02 \\
    rscvn & RS CVn binary & 1593 & 0.67 & 0.35 & 0.92 & 0.59 \\
    ceph & classical Cepheid variable star & 1573 & 0.90 & 0.63 & 0.95 & 0.68 \\
    saw & sawtooth (p) & 1314 & 0.83 & 0.17 & 0.40 & 0.01 \\
    srv & semiregular variable star & 950 & 0.71 & 0.05 & 0.72 & 0.61 \\
    agn & active galactic nucleus & 596 & 0.52 & 0.24 & 0.93 & 0.68 \\
    rrd & RR Lyr d star & 539 & 0.52 & 0.28 & 0.66 & 0.43 \\
    ceph2 & Population II Cepheid variable star & 281 & 0.50 & 0.08 & -- & 0.0 \\
    mir & Mira variable star & 248 & 0.65 & 0.50 & 0.59 & 0.45 \\
    ext & Spatially extended source (p) & 154 & -- & 0.0 & 0.50 & 0.07 \\
    wvir & W Vir variable star & 77 & 0.33 & 0.20 & 1.0 & 0.40 \\
    \\
    wp & wrong period (p) & 1694 & -- & 0.0 & -- & 0.0 \\
    el & ellipsoidal (p) & 931 & -- & 0.0 & 0.0 & 0.0 \\
    blend & blended sources (p) & 360 & -- & 0.0 & -- & 0.0 \\
    mp & multi-periodic (p) & 257 & -- & 0.0 & -- & 0.0 \\
    dp & double period (p) & 238 & -- & 0.0 & -- & 0.0 \\
    hp & half period (p) & 237 & -- & 0.0 & -- & 0.0 \\
    bright & nearby bright star (p) & 106 & -- & 0.0 & -- & 0.0 \\
    blher & BL Her star & 95 & -- & 0.0 & 0.0 & 0.0 \\
    rrblz & RR Lyrae (Blazhko effect\footnote{\citet{blazhko1907}}) & 85 & -- & 0.0 & -- & 0.0 \\
    dip & dipping (p) & 64 & -- & 0.0 & -- & 0.0 \\
  \hline
\hline
\end{tabular}
\tablecomments{Here (p) denotes phenomenological classifications whose classifiers receive only the subset of features specified in Table \ref{tab:phenom_features}.}
\end{center}
\end{table*}

\subsection{Inference}
We used the trained models\footnote{\texttt{trained\_dnn\_models.zip} and \texttt{trained\_xgb\_models.zip} at \url{https://zenodo.org/doi/10.5281/zenodo.8410825}} to perform inference on 77 ZTF fields, including the original 20 studied in \paperone\ (see Figure 4 of that work). These fields represent different parts of the sky (e.g. in and out of the Galactic plane, toward and away from the bulge). To expand the sample to 77, we initially added a field on each ``side'' of the original 20. For example, we added fields 295 and 298 to the original pair of 296 and 297. We then added fields having  immediately lower and higher declinations compared to the existing collection. Overall, the 77 fields yielded 209,991,147 sets of features and classifications, each corresponding to a ZTF light curve. The full collection of DNN and XGB classifications is available on Zenodo;\footnote{\texttt{predictions\_dnn\_xgb\_***\_fields.zip} file at \url{https://zenodo.org/doi/10.5281/zenodo.8410825}} a partial table of the first 10 rows of predictions for field 487 is shown in Table \ref{tab:preds-partial}.

\begin{sidewaystable}
\begin{center}
\caption{Partial Table of DNN and XGB Predictions for ZTF Field 487.}
\label{tab:preds-partial}
\begin{tabular}{lrrrrrrrrrrrrrr}
\hline
\hline
\_id & Gaia\_EDR3\_\_\_id & AllWISE\_\_\_id & PS1\_DR1\_\_\_id & ra & dec & period & field & ccd & quad & filter & e\_dnn & ... & saw\_xgb & ... \\
\hline
\hline
10487361000000 & 4285766690080980352 & 2825106001351047680 & 115652827469136144 & 282.747 & 6.380 & 0.025 & 487 & 10 & 1 & 1 & 0.01 & ... & 0.0 & ... \\
10487361000001 & 4285763941301551232 & 0 & 115652824753267152 & 282.475 & 6.381 & 0.021 & 487 & 10 & 1 & 1 & 0.01 & ... & 0.0 & ... \\
10487361000002 & 4285763937000857472 & 0 & 115652824816236864 & 282.482 & 6.380 & 0.384 & 487 & 10 & 1 & 1 & 0.00 & ... & 0.0 & ... \\
10487361000003 & 4285809433595186176 & 2825106001351042560 & 115652823552896816 & 282.355 & 6.380 & 0.024 & 487 & 10 & 1 & 1 & 0.08 & ... & 0.0 & ... \\
10487361000004 & 4285765453130410112 & 2825106001351046656 & 115652826992225120 & 282.699 & 6.379 & 0.021 & 487 & 10 & 1 & 1 & 0.02 & ... & 0.0 & ... \\
10487361000005 & 4285777719557768576 & 2825106001351048704 & 115652829454493488 & 282.945 & 6.378 & 23.731 & 487 & 10 & 1 & 1 & 0.01 & ... & 0.0 & ... \\
10487361000006 & 4285810052070487680 & 2825106001351042560 & 115652823125995488 & 282.313 & 6.379 & 0.057 & 487 & 10 & 1 & 1 & 0.01 & ... & 0.0 & ... \\
10487361000007 & 4285763185387287936 & 2825106001351044096 & 115652825352683952 & 282.535 & 6.378 & 118.653 & 487 & 10 & 1 & 1 & 0.02 & ... & 0.0 & ... \\
10487361000008 & 4285763941301548032 & 2825106001351044608 & 115652824833284160 & 282.483 & 6.378 & 0.023 & 487 & 10 & 1 & 1 & 0.02 & ... & 0.0 & ... \\
10487361000009 & 4285755076490209280 & 2825106001351047680 & 115652828372002672 & 282.837 & 6.377 & 0.021 & 487 & 10 & 1 & 1 & 0.01 & ... & 0.0 & ... \\
... & ... & ... & ... & ... & ... & ... & ... & ... & ... & ... & ... & ... & ... & ... \\
\hline
\hline
\end{tabular}
\end{center}
\tablecomments{Values of 0 for Gaia, AllWISE, and Pan-STARRS IDs indicate the lack of a cross-matched source in that survey. For brevity, coordinates and period have been rounded to three decimal places. In addition, only one prediction column per algorithm is listed.}
\end{sidewaystable}

\begin{figure*}
    \centering
    \includegraphics[scale=0.5]{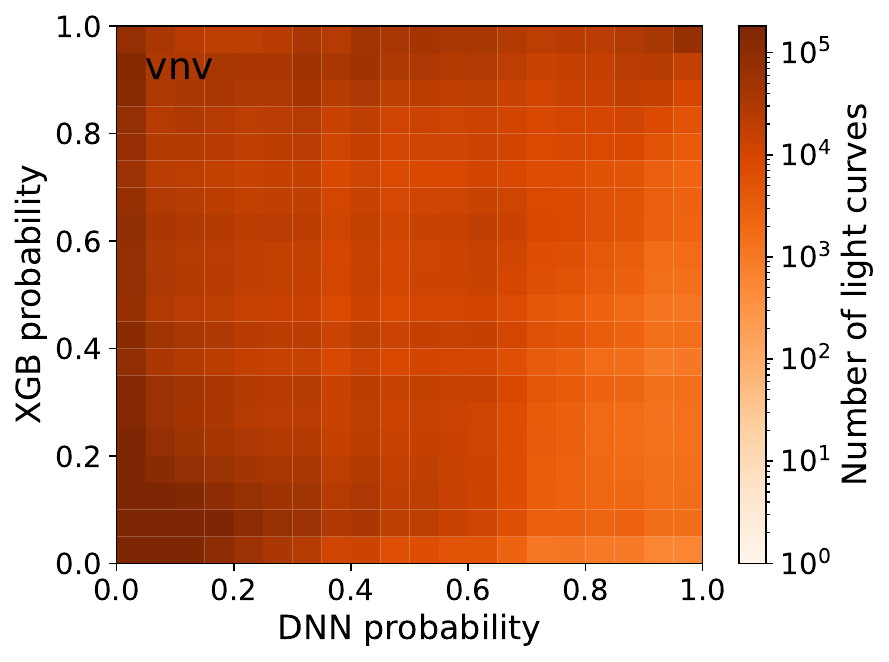}
    \includegraphics[scale=0.5]{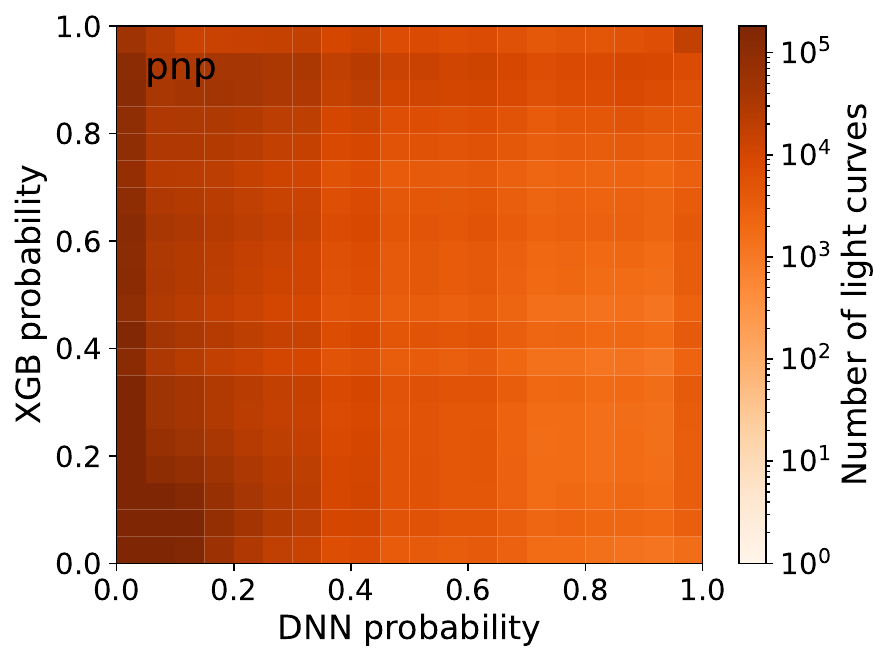}
    \includegraphics[scale=0.5]{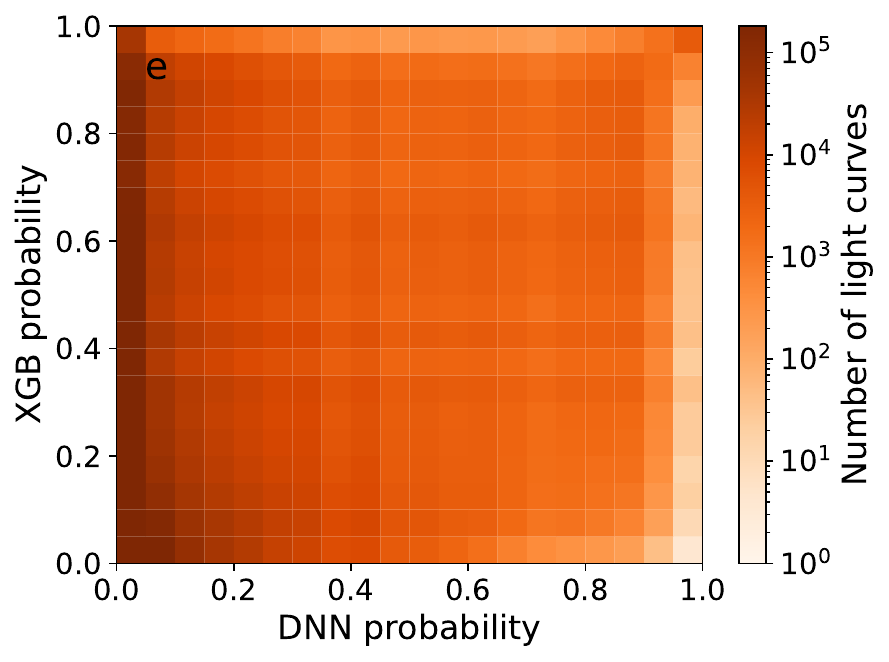}
    \includegraphics[scale=0.5]{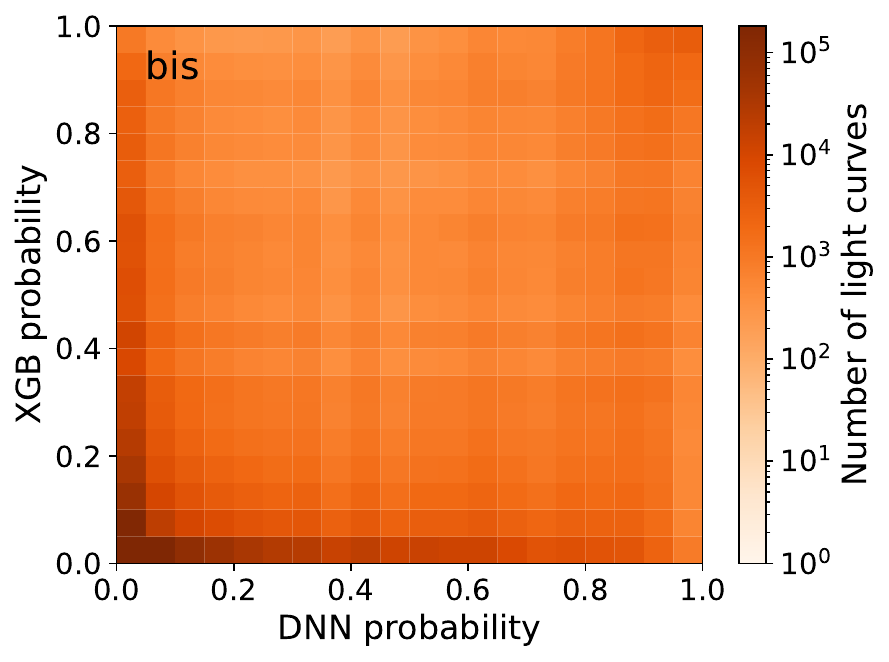}
    \caption{Heatmaps of XGB and DNN classification probabilities (fields 487, 563, and 777) for the four classes with the most positive training examples.}
    \label{fig:heatmaps_top4}
\end{figure*}

\begin{figure*}
    \centering
    \includegraphics[scale=0.5]{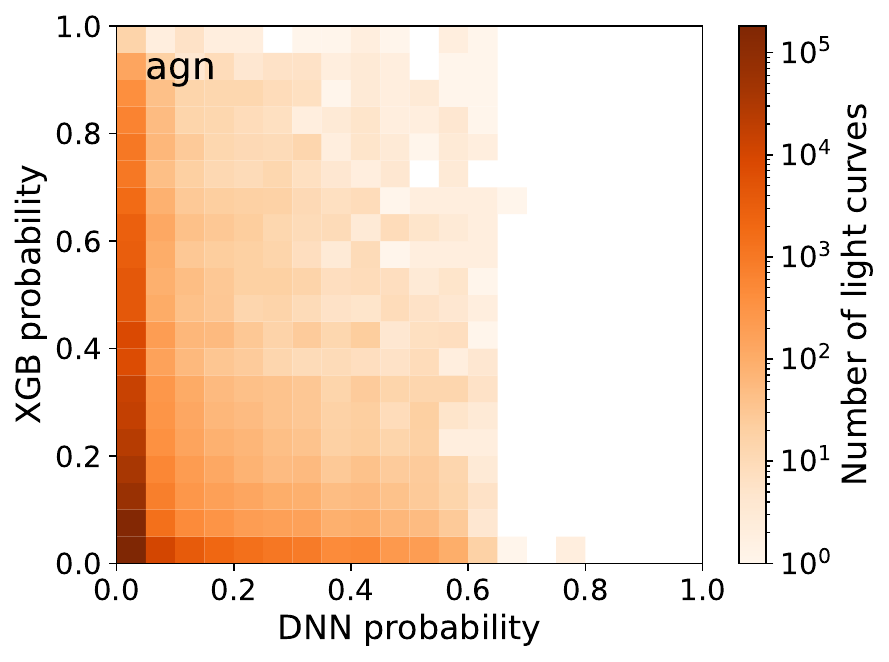}
    \includegraphics[scale=0.5]{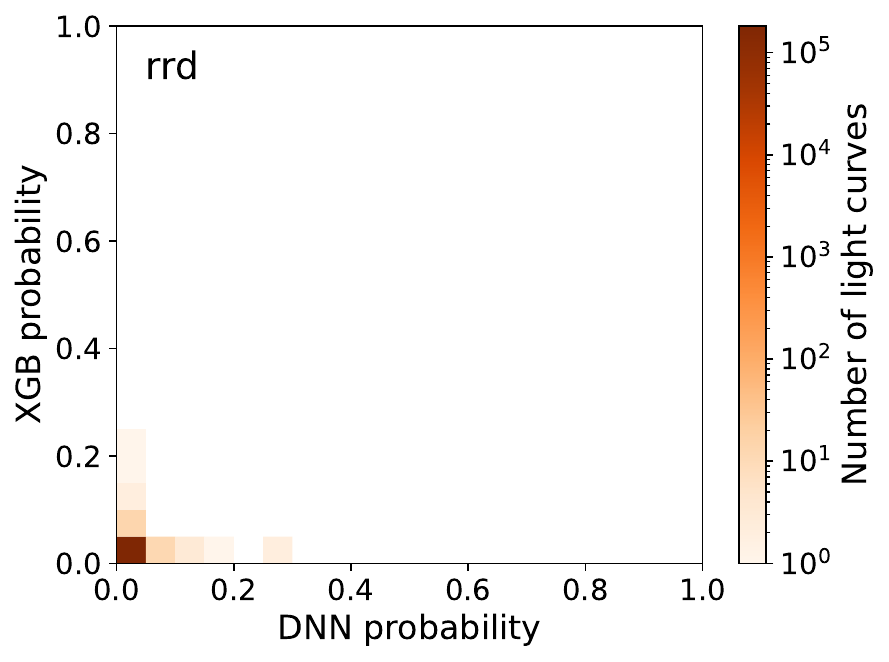}
    \includegraphics[scale=0.5]{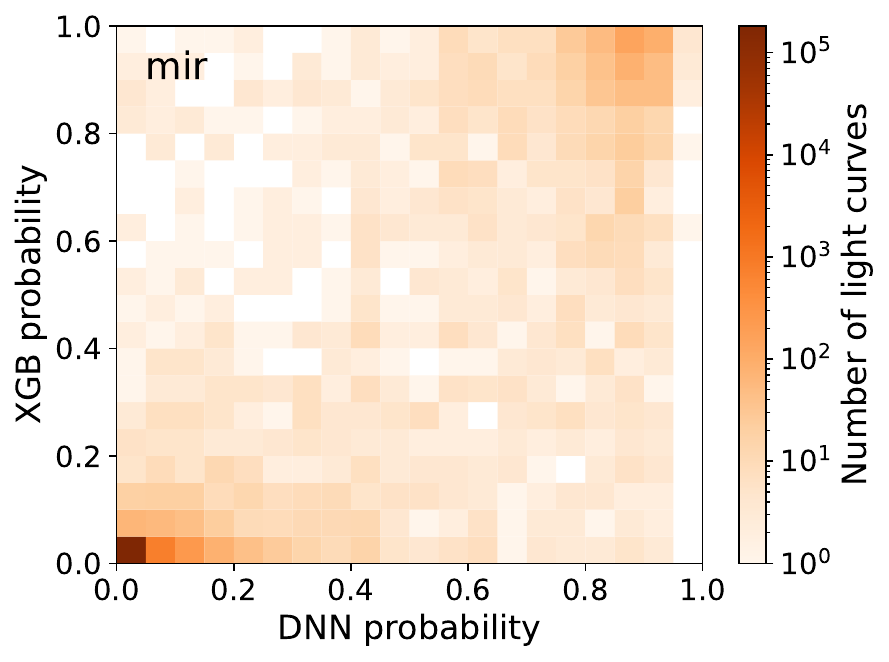}
    \includegraphics[scale=0.5]{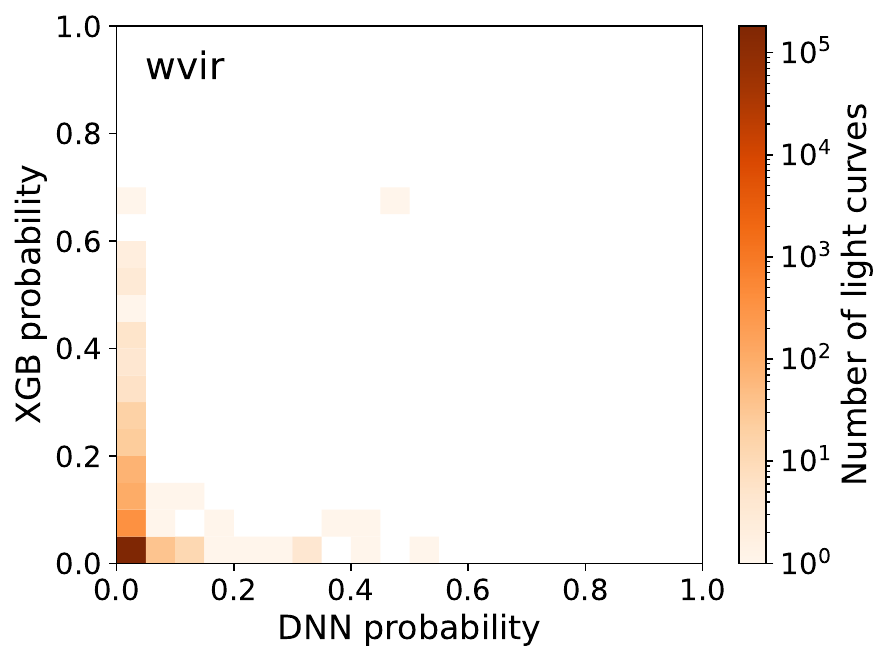}
    \caption{Heatmaps of XGB and DNN classification probabilities (fields 487, 563, and 777) for the four classes with the fewest positive training examples.}
    \label{fig:heatmaps_bottom4}
\end{figure*}

Figures \ref{fig:heatmaps_top4} and \ref{fig:heatmaps_bottom4} show heatmaps of predicted DNN and XGB classification probabilities for sources in fields 487, 563, and 777 (containing 18,184,402 light curves). Figure \ref{fig:heatmaps_top4} plots the binned probabilities for the top four most represented classes in the training set, while Figure \ref{fig:heatmaps_bottom4} does the same for the four least represented classes that have nonzero precision/recall scores for both algorithms.

\section{Discussion}
\label{sec:discussion}

\subsection{Training Results}

\subsubsection{Precision and Recall}
While DNN and XGB performance varied from class to class, for both algorithms there is an association between the number of positive examples and precision/recall scores. With some exceptions, we report high precision and recall scores for classes at or above the median number of positive examples ($\sim$ 3000 light curves, or $\sim$ 2\% of all learning set light curves). This pattern is understandable since a classifier that is given more positive examples may generalize better than one trained on a smaller sample, as long as there remains a sufficient amount of negative examples for comparison. Among the classes with the most positive examples even the \texttt{vnv} classifier is given $\sim 25\%$ negative examples during training, avoiding a major class imbalance.

On the other end of the range, several classifiers were given fewer than 1000 positive examples for training, resulting in a $>99\%$ rate of negative examples. While this does not guarantee poorer classifier performance, it may make it more difficult for a classifier to learn and generalize. Many of the least represented classes achieve greater precision than recall, a result of exposing each classifier on the entire training set rather than weighting classes based on representation. While the ideal classifier achieves both high precision and recall scores, we chose to train in a way that favors precision over recall to minimize the number of false-positive classifications.
 
An additional factor that may contribute to the lower scores of some classifiers is the consistency of manual labels across the entire training set. The \texttt{sin} classifier offers a useful example: sinusoidal light-curve phenomenology should be readily identifiable based on the features input to the classifier. However, the \texttt{sinusoidal} label may not have been consistently applied, as human experts focused on their specific area of ontological expertise. As a result, the training set may contain sinusoidal light curves that are missing the appropriate label, reducing the \texttt{sin} classifier's performance. Finally, the intrinsic similarity of certain classes (e.g. contact binaries labeled as \texttt{wuma} and interacting \texttt{blyr} binaries, along with their \texttt{ew} and \texttt{eb} phenomenologies) adds further difficulty to the training of some classifiers.

\subsubsection{DNN versus XGB Training Scores}
The upper right corners of the scatter plots in Figure \ref{fig:precision_recall_scatter} show that for the classes containing higher numbers of positive examples the precision and recall scores are both high and comparable between DNN and XGB. There are also some classifiers trained on a more moderate number of positive examples that achieve high precision/recall. As the number of positive examples decreases, XGB tends to outperform DNN in precision scores. XGB also achieves higher recall for more classes than DNN, but the discrepancy is smaller than that for the algorithms' precision.

The general trend of higher precision for XGB classifiers may reflect a benefit of the algorithm's boosting approach as described in Section \ref{subsubsec:xgboost}, despite the lack of the \texttt{dmdt} feature input to DNN classifiers. The boosting approach does increase the chance of overfitting, which is quantified by Figure \ref{fig:test_train_scores_diff}. In these plots, differences between test and training set precision/recall scores that are near zero suggest little overfitting. For more negative differences on the plots, the model's superior performance on the training set indicates a potential overfit. The majority of DNN and XGB classifiers have test/training score differences near zero. However, some classifiers show signs of overfitting, especially the \texttt{sin}, \texttt{blend}, \texttt{ceph2}, and \texttt{mir} DNN classifiers and the \texttt{emsms}, \texttt{blyr}, \texttt{sin}, \texttt{rscvn}, \texttt{agn}, \texttt{rrd}, \texttt{ceph2}, and \texttt{mir} XGB classifiers. These classifiers tend to be trained on fewer positive examples than classifiers that show minimal overfitting.

\subsubsection{Feature Importance}
According to Figure \ref{fig:most_important_features_top3}, the feature most frequently having top three importance among XGB classifiers is \texttt{period\_ELS\_ECE\_EAOV}. This quantity is the light curve's period determined by the nested approach described in Section \ref{subsec:period_finding}. This feature was only input to ontological classifiers, and its high importance is consistent with the well-defined period ranges known to be indicative of many ontological classes in \scope. 

The next feature among the top three, \texttt{f1\_power\_ELS\_ECE\_EAOV}, is the normalized $\chi^{2}$ value associated with the best-fitting Fourier series to the light curve. A value close to 1 means that the fourth-order Fourier series fit best, while a value close to zero indicates that the zeroth order (constant value) provided the best fit. Given the variable and periodic nature of most \scope\ classes, it is unsurprising that this indicator of variability was highly important in many cases. The ability to use a parallax value to convert apparent to absolute magnitudes supports the inclusion of \texttt{Gaia\_EDR3\_\_parallax} among the top features in Figure \ref{fig:most_important_features_top3}. We note that even negative parallax values retain a meaningful connection to an object's distance despite larger uncertainties \citep[e.g.][]{luri2018}, and thus their inclusion in our feature set is warranted.

Rounding out the top five features are \texttt{Gaia\_EDR3\_phot\_bp\_rp\_excess\_factor} and \texttt{inv\_vonneumannratio} . The von Neumann ratio computes the ratio between the correlated variance and the variance, and it is thus sensitive to variability. The $BP - RP$ excess factor evaluates the flux ratio $(I_{BP} + I_{RP}) / {I_{G}}$. This statistic, originally intended as a measure of photometric quality, may also serve as a proxy for color, which often delineates different astronomical objects.

Our feature importance results show similarities with those obtained by \citet{richards2012_asas} when training a random forest classifier on ASAS sources. Both studies found the period (or fundamental oscillation frequency) to be the most important feature. Additionally, both list the Stetson $J$ coefficient and skew among the top 10 most important features. Going forward, results like these may be used to reduce the number of features required for reliable classifications in the future. This reduction would expedite future classification projects, potentially to the point of enabling real-time classification based on a small number of important features.  

\subsection{Inference Results}
The heatmaps in Figures \ref{fig:heatmaps_top4} and \ref{fig:heatmaps_bottom4} show a spread of DNN/XGB classification probabilities across the 2D space for a combination of fields 487, 563, and 777 (18,184,402 light curves). Perfect agreement between algorithms would be achieved if only the diagonals on these plots contained light curves. Multiple patterns are discernible from the heatmaps in Figure \ref{fig:heatmaps_top4}. The largest number of sources in each heatmap is in the lower left corner, where both algorithms share a probability $< 0.05$. This indicates agreement among both algorithms that many light curves are unlikely to be classified with each label. For many of the visualized classifiers, there is also a concentration of light curves sharing high DNN/XGB confidence in each classification. This agreement appears as the darker squares in the upper right corners of some panels. In other cases (such as the \texttt{srv} classifier), very few probabilities approach unity, but a correlation is visible between DNN/XGB probabilities.

Especially for the \texttt{pnp}, \texttt{e}, and \texttt{bis} heatmaps, there are single columns or rows indicating a wide range of one algorithm's probabilities paired with a narrow range from the other. For example, the \texttt{pnp} heatmap shows a column with many light curves having a wide range of XGB probabilities but DNN probabilities between 0 and 0.05. Similarly, there is a row in the \texttt{bis} heatmap showing many DNN probabilities for a narrow range of XGB probabilities. While these features indicate some level of disagreement between the regression performed by the algorithms, the mapping of probabilities to dichotomous classifications using a threshold (probability $> 0.7$) results in the high precision/recall scores for all four classifiers shown in Figure \ref{fig:heatmaps_top4}.

Figure \ref{fig:heatmaps_bottom4} shows some familiar patterns from Figure \ref{fig:heatmaps_top4} and some new ones for the four least represented labels. Again, the lower left corner contains the most light curves, indicating agreement between DNN and XGB for many near-zero probabilities. Some panels also show columns or rows indicating a wide range of probabilities from one algorithm paired with a narrow range from the other algorithm (especially for the \texttt{agn} heatmap). 

In the case of the \texttt{mir} heatmap, a pattern of agreement is visible between DNN and XGB, as indicated by the high density of light curves in plot's upper right and lower left corners. The other heatmaps in Figure \ref{fig:heatmaps_bottom4} do not have any light curves along their top and right edges, showing that some classifiers do not yield probabilities near unity. For \texttt{rrd} and \texttt{wvir}, very few probabilities are $> 0.5$. This highlights the importance of considering not only absolute probabilities when evaluating \scope\ predictions but also the probabilities relative to each classifier's distribution. For example, one might select candidate W Vir variables by considering light curves having a top percentile of DNN and XGB probabilities, even if they are not high on an absolute scale.

\subsubsection{DNN versus XGB Proabilities}
Although inference results for unclassified light curves cannot be compared with ground truth in the same way as our training data, there are still useful insights to be learned from the collection of predictions. For example, for fields 487, 563, and 777 we study the agreement among DNN and XGB classifications on a per-classifier basis, considering the same probability threshold of 0.7 we used for algorithm training. We find that across all classifiers having nonzero recall scores for both ML algorithms, an average of 99\% of light curves have both DNN and XGB probabilities either greater than or less than 0.7 (i.e. not conflicting given this threshold). The range of this agreement is between 85\% and 100\% depending on the classifier.

However, the classifiers showing the greatest agreement using the above method typically score so highly because few or no light curves are classified with probabilities greater than 0.7. For these classifiers, we therefore adjust the test by iteratively decreasing the DNN and XGB probability thresholds independently until at least 1000 light-curve probabilities were above each threshold. This produces little change in the agreement fractions between DNN and XGB for all classifiers, except for a very slight ($\sim 0.003\%$) decrease in the maximum fraction of agreeing classifications.

While the above results imply strong agreement between DNN and XGB, they remain biased by the fact that, especially for the more specific ontological classifications, most light curves will have probabilities close to zero, indicating a nonclassification.
This is visualized by the dark squares in the lower left corners of each heatmap, especially those in Figure \ref{fig:heatmaps_bottom4}. 
While agreement on nonclassifications is important, an additional useful test of agreement is to consider only the top $N$ light-curve probabilities from each classifier.

For this test of high-confidence classification agreement between DNN and XGB, we analyze a different number of $N$ light curves for each classifier in order to account for differences in the classifiers' levels of specificity. For example, we expect far more light curves to be classified with high probabilities from the \texttt{vnv} classifier than we do from the \texttt{mir} classifier.

We therefore use the sum of all light-curve probabilities from a given classifier as a rough proxy for the relative frequency of that class. This produces a number between 0 and the 18,184,402 light curves across the three fields, thus weighting the value $N$ on a per-class basis. For the classifiers mentioned above, we consider the top 3,691,883 \texttt{vnv} light curves and the top 1,710 \texttt{mir} light curves. Across all classifiers, we find that an average of 36\% of top $N$ classifications agree between DNN and XGB, with a range between 1\% and 85\%. While this analysis does not offer insight into the ground truth of these classifications, it shows that there is a mix of agreement and disagreement between the two algorithms. Areas of agreement among both algorithms correspond to the highest-confidence classifications in the sample, while areas of disagreement represent interesting conflicts that may indicate a preferred ML algorithm for that class or anomalous light curves.

Finally, we study the connection between hierarchical classifications. \scope\ classifiers are trained independently, and the predicted classifications are not influenced by each other. To explore the results of this approach, we examine the immediate subclasses of the \texttt{vnv} label, \texttt{pnp} and \texttt{i}. Among light curves in fields 487, 563, and 777 with \texttt{pnp} probability $> 0.7$, 74\% also have \texttt{vnv} probability $> 0.7$ for DNN. The same is true for 94\% of XGB light curves. Using the same probability threshold, 91\% of DNN \texttt{i} light curves are also \texttt{vnv} (92\% for XGB). These results suggest that a logical hierarchy generally persists among these independent classifications, and we see the most consistency for the XGB algorithm.

\section{Conclusion}
\label{sec:conclusion}
In this paper, we have reported the training and inference results for the open-source ZTF Source Classification Project, which trains dichotomous ML classifiers on ZTF data. The two algorithms we used, a neural network and XGBoost, achieved comparable precision and recall scores for several well-represented classes. As the number of positive training examples decreased, the classifiers displayed more noticeable differences in performance. In particular, XGB often scored higher in precision than DNN as the number of positive examples decreased. Recall scores were more comparable between the algorithms. 

We used the XGB algorithm to determine feature importance across classifiers, finding that the light-curve periods were most often among the features of highest importance. This feature was followed in importance by a quantity encoding the order of the best-fitting Fourier series to the light curve, the Gaia EDR3 parallax and $BP - RP$ excess factor, and the inverse von Neumann ratio. Future work could maximize computational efficiency by reducing the number of included features to the minimal amount required for reliable results.

We reported classification predictions for 209,991,147 light curves using 34 dichotomous classifiers. This catalog of DNN and XGB classification predictions, as well as the training set, is available electronically on Zenodo. The computational demands of running inference on all ZTF fields limits this paper's reported predictions to these 77 fields, which represent a wide range of regions on the sky. This variable source catalog will continue to grow as additional ZTF fields are run through the \scope\ workflow.

Future time-series ML work may streamline the feature generation process, reducing the resources required to classify a larger collection of light curves. The incorporation of additional ML algorithms may lead to improved performance on a broader variety of classes and numbers of positive examples. Finally, upcoming sources of new data will support future time-domain studies: the Legacy Survey of Space and Time at Rubin Observatory \citep{ivezic2019_lsst} will provide time-series data for nearly an order of magnitude more sources than ZTF, and the NEO Surveyor Mission \citep{mainzer2023_neosurveyor} will similarly succeed the Wide-field Infrared Survey Explorer in the mid-IR. The \scope\ code is readily adaptable to data with different cadences and bands, and we look forward to the continued contributions this project can make to time-domain astronomy.

\begin{acknowledgments}
We are grateful to the referee for providing helpful comments that strengthened the paper. B.F.H. and M.W.C. acknowledge support from the National Science Foundation with grant Nos. PHY-2308862 and PHY-2117997. This work used Expanse at the San Diego Supercomputer Cluster through
allocation AST200029, ``Towards a complete catalog of variable
sources to support efficient searches for compact binary mergers and
their products,'' from the Advanced Cyberinfrastructure Coordination
Ecosystem: Services \& Support (ACCESS) program, which is supported by
National Science Foundation grant Nos. 2138259, 2138286, 2138307,
2137603, and 2138296.

Based on observations obtained with the Samuel Oschin
48-inch telescope and the 60-inch telescope at the Palomar Observatory as part
of the Zwicky Transient Facility project. ZTF is supported by the
National Science Foundation under grant Nos. AST-1440341 and
AST-2034437 and a collaboration including current partners Caltech,
IPAC, the Weizmann Institute of Science, the Oskar Klein Center at
Stockholm University, the University of Maryland, Deutsches
Elektronen-Synchrotron and Humboldt University, the TANGO Consortium
of Taiwan, the University of Wisconsin at Milwaukee, Trinity College
Dublin, Lawrence Livermore National Laboratories, IN2P3, University of
Warwick, Ruhr University Bochum, and Northwestern University and former
partners the University of Washington, Los Alamos National
Laboratories, and Lawrence Berkeley National Laboratories. Operations
are conducted by COO, IPAC, and UW.

The Gordon and Betty Moore Foundation, through both the Data-Driven
Investigator Program and a dedicated grant, provided critical funding
for SkyPortal.

The data are available on Zenodo under an open-source 
Creative Commons Attribution license: 
\dataset[doi:10.5281/zenodo.8410825]{https://doi.org/10.5281/zenodo.8410825}.

\facilities{PO:1.2m \citep{bellm2019_ztf, graham2019_ztf, dekany2020_ztf, masci2019_ztf}, Gaia \citep{gaia2016, gaia2020_edr3}, WISE \citep{wright2010_wise, cutri2021_allwise}, PS1 \citep{kaiser2002_panstarrs, chambers2016_panstarrs1}}

\software{\texttt{scope-ml} (\url{https://github.com/ZwickyTransientFacility/scope-ml}), \texttt{cesium} \citep{naul2016_cesium}, \texttt{jupyter} \citep{granger2021_jupyter}, \texttt{kowalski} \citep{duev2019}, \texttt{matplotlib} \citep{hunter2007_matplotlib}, \texttt{numpy} \citep{oliphant2006_numpy1, vanderwalt2011_numpy2}, \texttt{pandas} \citep{mckinney2010_pandas}, \texttt{penquins} (\url{https://github.com/dmitryduev/penquins}), \texttt{periodfind} (\url{https://github.com/ZwickyTransientFacility/periodfind}), \texttt{scikit-learn} \citep{pedregosa2011_scikit-learn}, \texttt{SkyPortal} \citep{vanderwalt2019_skyportal, coughlin2023_skyportal}, \texttt{tdtax} (\url{https://github.com/profjsb/timedomain-taxonomy}), \texttt{tensorflow} \citep{tensorflow2015-whitepaper}, \texttt{wandb} \citep{wandb}, \texttt{XGBoost} \citep{chen2016_xgboost}}

\end{acknowledgments}


\bibliographystyle{aasjournal}
\bibliography{references}

\label{lastpage}
\end{document}